\documentclass[12pt]{article}  
\usepackage{makeidx}
\usepackage{graphicx}
\usepackage{amsmath} 
\usepackage{verbatim}   
\usepackage{rotating} 
\usepackage{fancyvrb}
\usepackage{float}
\usepackage[semicolon, authoryear]{natbib} 
\usepackage{url}  
\usepackage{algorithm}
\usepackage{algpseudocode} 
\usepackage{algorithmicx}
\usepackage{fancyhdr} 
\usepackage{tabularx,booktabs}
\usepackage{tikz}
\usetikzlibrary{positioning}
\usetikzlibrary{calc}

\newcommand{\mysection}[1]{ \vspace{24pt} \begin{center} \textbf{#1} \end{center}}

\begin{document}
	
  \baselineskip24pt
  
\title{\bf How to Tell When a Result Will Replicate: Significance and Replication in Distributional Null Hypothesis Tests}
\author{Fintan Costello \\ School of Computer Science and Informatics,\\ University College Dublin\\
	and \\
	Paul Watts \\ Department of Physics,\\National University of
	Ireland  Maynooth\\ }

 \thispagestyle{empty}

 \date{}
 \maketitle
 
 \noindent Submitted to Meta-Psychology. Participate in open peer review by sending an email to open.peer.reviewer@gmail.com. The full editorial process of all articles under review at Meta-Psychology can be found following this link:  \url{https://tinyurl.com/mp-submissions}
 You will find this preprint by searching for the first author's name.
 
 \subsection*{Author Note}
 Fintan Costello: Orcid ID \url{https://orcid.org/0000-0002-3953-7863}.\\
 Paul Watts: Orcid ID \url{https://orcid.org/0000-0002-9440-4387}.
 
 \noindent All data, analysis and visualisation code is publicly available on the OSF repository at \url{https://osf.io/ep86y}.
  Corresponding author: Fintan Costello (fintan.costello@ucd.ie), School of Computer Science, University College Dublin, Belfield, Dublin 4, Ireland.

\newpage
	
	\bigskip  	
	\mysection{Abstract} 
   There is a well-known problem in Null Hypothesis Significance Testing: many statistically significant results fail to replicate in subsequent experiments.  We show that this problem arises because standard `point-form null' significance tests consider only within-experiment but ignore between-experiment variation, and so systematically underestimate the degree of random variation in results.  We give an extension to standard significance testing that addresses this problem by analysing both within- and between-experiment  variation.  This `distributional null' approach does not underestimate  experimental variability and so is not overconfident in identifying significance; because this approach addresses between-experiment variation, it gives mathematically coherent estimates for the probability of replication of significant results.   Using a large-scale replication dataset (the first `Many Labs' project), we show that many experimental results that appear statistically significant in standard tests are in fact consistent with random variation when both within- and between-experiment variation are taken into account in this approach.   Further, grouping experiments in this dataset into `predictor-target' pairs we show that the predicted replication probabilities for target experiments produced in this approach (given predictor experiment results and the sample sizes of the two experiments) are strongly correlated with observed replication rates.   Distributional null hypothesis testing thus gives researchers a statistical tool for identifying statistically significant and reliably replicable results.
	
 	{\it Keywords:  Hypothesis Testing; Replication; Significance }
	\vfill

\newpage

	\newpage

 Confronted with a surprising experimental result and wondering whether it may represent a meaningful discovery, a researcher must ask: could this result simply be a consequence of random variation and no real effect?  Such questions are especially important for discovery in domains of high variability 
 (psychology, neuroscience, medicine, genetics, and so on), where random variation can easily produce apparently surprising results.  Since by definition a surprising result has not been hypothesised \textit{a priori}, such questions cannot be answered by tests which compare hypotheses with distinct \textit{a priori} means \citep[e.g. Neyman-Pearson hypothesis testing or informed Bayesian model comparison;][]{christensen2005testing,berger1996intrinsic,gonen2005bayesian}.    The standard way to answer such questions is via Fisher's evidential  significance test,  which assesses the degree to which a  result is unexpected  under the null statistical model of random variation but no real effect (and so gives evidence against that model) operationalised as the probability $p$ of a result as or more extreme arising under that model.  
 
 While the logic of evidential significance testing is compelling (a result is surprising when it is extreme, relative to our expectations; a low $p$-value tells us that our result is  extreme, relative to the null model) it has become increasingly clear that significance testing, at least as it is typically used, is fundamentally flawed.   Perhaps the most important problem with standard significance testing involves  replication and the reliability of results, with standard significance testing  being overconfident in identifying nonsensical effects, such as telepathy or precognition, as real \citep[parapsychology `offers a truly alarming massive case study of how statistics can
 mislead and be misused':][]{diaconis1991replication}, and more generally, with systematic research showing that  many statistically significant experimental results in psychology and related fields do not occur reliably in replications \citep[e.g.][]{camerer2018evaluating,open2015estimating,klein2018many,klein2014investigating}, results seen as `reflecting an unprecedented level of doubt among practitioners about the reliability of research findings in the field' \citep{pashler2012editors}.   These flaws are also evident in well-known effects of sample size on significance (the observation that the probability of getting a statistically significant result in the standard approach increases with sample size, irrespective of the presence or absence of a true effect; to quote \citet{thompson1998praise}: `Statistical testing becomes a tautological search for enough participants to achieve statistical significance. If we fail to reject, it is only because we've been too lazy to drag in enough participants')  and in the fact that in standard significance tests, null hypotheses are always false (and to quote Cohen, 2016: `if the null hypothesis is always false, what's the big deal about rejecting it?').   Faced with these problems and others \citep[primarily related to hypothesis comparison and optional stopping, e.g.][]{wagenmakers2018bayesian,wagenmakers2007practical} there have been  increasingly widespread calls for the wholesale abandonment of the significance testing approach \citep[e.g.][]{BlakeleyAbandon2019,amrhein2018remove,wagenmakers2018bayesian,hunter1997needed,carver1978case}.

   We see these problems as arising from a basic gap in the standard approach to  significance testing: the fact that it considers only within-experiment random variation in estimation, while ignoring between-experiment random variation. In the language of measurement analysis in physics, the standard approach considers the \textit{precision} of the experiment as a measurement tool (the degree of random variation in measurements within this experiment, a factor which depends on sample size) but ignores the \textit{accuracy} of the experiment as a tool \citep[the degree to which measurements in this experiment agree with those obtained from other experiments measuring the same effect; e.g][]{mandel2012statistical}.  This gap means that the standard approach systematically underestimates the degree of random variation or error affecting experimental results, and so is overconfident in identifying results as significant.  We address this by describing a model of significance testing that includes both within- and between-experiment random variation; the chance of statistical significance in this `distributional' model does not rise with sample size irrespective of any real effect; this approach does not overestimate the random variation affecting experimental results and so is not overconfident in identifying significance;
    importantly, this model gives coherent estimates for the probability of replication of significant results in future experiments.  
   
   The idea of between-experiment variation is, of course, well known, and is referred to as `heterogeneity' in the meta-analytic and systematic review literatures.  Researchers carrying out a systematic review of a set of studies on a given result will typically distinguish between clinical heterogeneity or diversity (between-experiment variation in the participants, interventions and outcomes studied), methodological heterogeneity (variation in study design,  measurement tools, and control of confounding factors), and outcome heterogeneity (variation in the precise effect being evaluated in each experiment) with outcome heterogeneity  arising  as a consequence of clinical or methodological differences between studies \citep[see e.g.][]{chandler2019cochrane}.   One strand of research in this area is causal, and focuses on `hidden moderators': particular differences in study design, participant population, or measured outcome that are hypothesised to produce the observed difference in experimental results \citep{kenny2019unappreciated,tackett2017s}.  Such hidden moderators are particularly plausible when considering the heterogeneity of psychological results across cultures: studies designed in a particular cultural setting may implicitly depend on some specific aspect of that culture,  with participants from a different cultural background giving obviously different responses.  A clear example of this arises in our analysis of results from the Many Labs $1$  project  \citep{klein2014investigating}, which ran studies replicating $16$ classic cognitive and social psychology results at $36$ different sites across the world (see below).   One of these studies examined the effect of `anchoring' on responses to numerical questions by asking participants to estimate the height of Mount Everest: where estimates given by participants at US sites  were typically quite far from the correct height,  participants at some international sites gave estimates that were only a few metres from the correct height (indicating a significant cross-cultural difference in participant knowledge about the actual height of Mount Everest).  
   
 In the `hidden moderator' approach, heterogeneity in experimental results arises as a consequence of one or two major and identifiable differences in experimental method or participant population: in this perspective heterogeneity can be controlled by careful design.  A contrasting `random-effects' approach sees heterogeneity as arising, not from one or two major differences between experiments, but instead from the combined effect of many small and random differences: from the Central Limit theorem, this combined effect will produce Normally distributed variability between experimental means, just as the combined effect of small and random differences between individual participants produces Normally distributed variability in participant responses.  This assumption of Normality allows us to use standard statistical methods to estimate between-experiment variance  \citep[see e.g.][]{hedges1983random,viechtbauer2005bias,veroniki2016methods}: in systematic reviews this type of random-effects analysis is used  to increase precision in our overall estimate of the size of an effect of interest by combining results from different studies, giving more weight to results with lower within-experiment variance.\footnote{Note that in presenting our approach we use the  within-experiment/between-experiment variance terminology, rather than distinguishing between `variance' and `heterogeneity'.  We make this terminological choice because, following this random-effects approach, we see `heterogeneity' as just another form of statistical variance, and to stress that between-experiment variance arises even when experiments are not obviously heterogeneous (do not have obvious differences in participant population or methodology).}
   
   Rather than using  between-experiment variance to increase the precision of our total estimate for an effect (parameter estimation via \textit{post-hoc} meta-analysis of existing results, as in systematic review), our approach uses between-experiment variance  to construct an \textit{a priori} null hypothesis designed to include both within- and between-experiment variance when assessing the statistical significance of new experimental results.  This distributional null hypothesis assumes,  not that the true effect is $0$ across all experiments (as in standard or `point-form' null hypothesis tests), but instead that the true effect is randomly distributed across experiments around a mean of $0$ (with a given level of between-experiment variance).  A statistically significant experimental result, in this distributional approach, is one which is inconsistent with an hypothesised mean of $0$ even under this assumption of between-experiment variance.   We can illustrate this point using a recent `W-Boson mass anomaly' in particle physics, arising from 2022 W-Boson mass measurements by the Collider Detector at Fermilab collaboration \citep{cdf2022high}.  These measurements were significantly higher than the Standard Model prediction for that particle mass and were  inconsistent with previous experimental results at a $p < 10^{-6}$  significance level in standard statistical tests.  This anomaly was seen as suggesting new physics beyond the Standard Model, such as new particles or interactions that modify the $W$-boson's mass \citep[e.g.][]{endo2022new,ghorbani2022w,du2023explaining,zhang2023explaining}.  Using our model of significance to analyse estimates of the $W$-Boson mass across a range of particle physics experiments \citep{costello2023underestimating} we found that these estimates were distributed with Normal between-experiment error: the anomalous CDF measurements were no longer statistically significant when the effect of this between-experiment random variation was taken into account.

   The model we describe mixes a form of statistical inference typically associated with the Frequentist approach (significance testing) with a form typically associated with the Bayesian approach (prior distributions over effect sizes):   in section $1$ we give a general sketch and overview of the model, positioning it relative to these two approaches.   In section $2$ we derive basic expressions for significance, statistical power and probability of replication in this model in the context of a $t$ test, assuming known between-experiment variation. In  section $3$ we give unbiased sample estimators for between-experiment variance of means (following a standard meta-analytic approach), give expressions for significance and replication in terms of those sample estimates, and extend these to give simple and practical expressions for significance and replication in terms of the standard $t$ distribution.  We also extend this approach to tests of correlation and linear regression, and  tests of contingency, and briefly compare this model to a well-known point-form approach to replication \citep{killeen2005alternative,killeen2007replication}.    In section $4$ we test this model against the first `Many Labs' replication dataset \citep{klein2014investigating}, showing that this model's assumption of between-experiment variation holds, that its judgements of statistical significance are many orders of magnitude more conservative than those obtained in the standard approach, and that its replication probabilities reliably and accurately predict observed rates of successful replication of individual results.  In section $5$ we compare this model to Bayesian approaches to hypothesis testing via the JZS Bayesian $t$ test \citep[e.g.][]{rouder2009Bayesian}, showing that this Bayesian approach gives results almost identical to those of a point-form significance test, at least in the Many Labs dataset.  In the final section we address some possible criticisms of this model of significance and replication, and argue that this distributional approach to evidential significance testing will help researchers identify experimental results which are both surprising (statistically significant) and amenable to further investigation (replicable).  Rather than abandoning significance testing as a way of identifying surprising and interesting results, researchers should consider using significance testing in a more general distributional form.

   \section{Sketch of the distributional model}
   
   We begin by sketching our model in very broad strokes, by placing it relative to standard Frequentist and Bayesian approaches to statistics.  This preliminary sketch necessarily ignores many important details: these are addressed in the full model presentation, below.   
    
   The basic distinction between  Frequentist and Bayesian approaches to statistical inference lies in the interpretation of probability:  the Frequentist sees probability as the limit of relative frequency in sampling, while the Bayesian sees probability as representing degree of belief or uncertainty in estimates.  This distinction means that  Frequentist statistical inference considers the distribution of measurements in samples, while Bayesian inference considers both this sampling distribution and additional prior distributions (with these priors reflecting  degrees of belief or uncertainty in parameters of the sampling distribution).
   For example: in a simple experiment estimating some unknown parameter $\mu$ from $N$ measurements and assuming Normal error with unknown variance $\sigma^2$, the standard Frequentist approach assumes that the sample mean $\overline{X}$ is  distributed as
   \[ \overline{X} \sim \mathcal{N}(\mu,\sigma^2/N)\]
   (the Normal error model).  Defining the test statistic
   \[t=\frac{|\overline{X}|}{S/\sqrt{N}}\]
    and taking the sample standard deviation $S$ to be an accurate estimate of the model standard deviation $\sigma$, then a given result will be statistically significant relative to the null hypothesis $H_0:\mu=0$ at level $\alpha$ in a two-sided test when
   \[ P(\overline{X} | H_0) = 2\Phi\left(\frac{-|\overline{X}|}{S/\sqrt{N}}\right)  < \alpha \]
  where $P(\overline{X} | H_0)$ represents the probability of observing a result equal to or more extreme than $\overline{X}$ given $H_0$ and $\Phi$ represents the cumulative Normal distribution.  This is the $Z$ test for significance, and is valid only when we can reasonably assume that $S$ accurately estimates $\sigma$ (an assumption which only holds for large $N$).   We can generalise this approach by considering the distribution of the test statistic $t$ under the Normal error model: for an experiment with $\nu$ degrees of freedom, this statistic has the cumulative $t$ distribution $T_\nu$, and a result will be statistically significant relative  $H_0:\mu=0$ at level $\alpha$ in a two-sided test when
  \[  2 T_\nu \left(\frac{-|\overline{X}|}{S/\sqrt{N}}\right)  < \alpha \]
This distribution $T_\nu$ approaches $\Phi$ as the  sample size $N$ increases (that is, as $S$ more accurately estimates $\sigma$).

   Where this `Frequentist' approach assumes a null hypothesis with $\mu = 0$, the Bayesian approach assumes a  prior distribution $\mu \sim \mathcal{N}(\mu_0,\sigma_0^2)$ with  $\mu_0$ representing a prior estimate for the parameter of interest and $\sigma_0^2$ representing prior uncertainty or degree of belief in that estimate.  Here our statistical model is
\begin{equation*} 
\overline{X}  \sim \mathcal{N}(\mu,S^2/N) \textit{\, where  \,}  \mu \sim \mathcal{N}(\mu_0,\sigma_0^2) \textit{\, }  
\end{equation*}
or simplifying by integrating out $\mu$ 
\begin{equation*} 
\overline{X}  \sim \mathcal{N}(\mu_0,\sigma^2/N+\sigma_0^2) 
\end{equation*}
and so taking $H_0:\mu_0=0$, if we assume that $S^2$  accurately estimates $\sigma^2$ (as before, a reasonable assumption for large $N$) then we have
  \[ P(\overline{X} | H_0) = 2\Phi\left(\frac{-|\overline{X}|}{\sqrt{S^2/N+\sigma_0^2}}\right)   \]
 Bayesians, however, are typically more interested in the  updated or posterior distribution 
\[\mu \sim \mathcal{N}(\mu_N,\sigma_N^2)\]
which is obtained by applying Bayes' theorem to the prior and the $N$ sample measurements.  This gives a probability distribution for hypotheses $H$ given data $\overline{X}$  of
\[ P(H|\overline{X})= P(\mu=m|\overline{X}) = \mathcal{N}(m|\mu_N,\sigma_N^2)\]
which, in the Bayesian perspective, describes how we should update our beliefs about  $\mu$ given the observed data $\overline{X}$.
   
 The standard Frequentist approach considers only within-experiment variance of the mean, $\sigma^2/N$, but ignores any between-experiment variance in means.  Our distributional model extends the Frequentist approach to include between-experiment variance of  means, which we write as $\sigma_0^2$.   This gives a  distribution for means of 
 \[\overline{X} \sim \mathcal{N}(\mu_0, \sigma^2/N+\sigma_0^2) \]
   just as in the Bayesian approach.  If we assume that $\sigma$ is accurately estimated by the sample variance $S$ and that $\sigma_0^2$ by is accurately estimated by the sample variance or heterogeneity of means across similar experiments, $S^2_0$,   we can assess statistical significance relative to both within- and between-experiment variance: a given result will be statistically significant relative to the null hypothesis $H_0: \mu_0=0$ when
   \[ P(\overline{X}|H_0) = 2\Phi\left(\frac{-|\overline{X}|}{\sqrt{S^2/N + S^2_0}}\right)  < \alpha \]
   Just as before, we can also consider the distribution of the test statistic here under our statistical model for $\overline{X}$: it turns out that a particular transformation of this statistic follows the $t$ distribution with $\nu$ degrees of freedom as before, and so we can also test for significance against   $T_\nu$ (see  the next section). 
  
   We can also produce a posterior distribution  $\mu \sim \mathcal{N}(m|\mu_N,\sigma_N^2)$ by applying Bayes' theorem as before, giving a distribution  for hypotheses given data of $P(H|\overline{X})$.   In this distributional approach, both directions of inference (`Frequentist' estimation of the probability of data given hypothesis $P(\overline{X}|H)$,  and `Bayesian' estimation of the probability of  hypotheses given  data $P(H|\overline{X})$) can be applied simultaneously, with the difference between Frequentist and Bayesian viewpoints arising in the interpretation of $\sigma_0^2$ (the Frequentist interpreting it as representing between-experiment variance of experimental means and the Bayesian as representing degree of prior belief or uncertainty about the population mean).    Given the same experimental result $\overline{X}$ and the same values for $\mu_0$ and $\sigma_0^2$, however interpreted, in this approach Frequentists and Bayesians will make the same judgement of statistical significance and will infer the same updated distribution for $\mu$.   
   
    Our position, then, is that Bayesian and Frequentist approaches to statistical inference differ in their interpretation of probability, but not in the analyses that can be applied to mathematical representations of probability in terms of statistical distributions.  It happens that analyses which estimate $P(\overline{X}|H)$ are primarily associated with the Frequentist viewpoint (most likely because estimates $P(H|\overline{X})$ are uninformative  under a point-form null hypothesis: when the probability density of $H$ is massed at a single point, $P(H|\overline{X})=P(H)$ necessarily holds) while those which estimate $P(H|\overline{X})$ are primarily associated with the Bayesian approach.\footnote{ Although some Bayesian proposals do estimate $P(\overline{X}|H)$, typically as part of a prior check of validity before Bayesian inference \cite[e.g.][]{dass2003unified,gabry2019visualization,van2021bayesian,schad2021toward}.}    Our distributional model, by contrast, makes use of both  $P(\overline{X}|H)$ and  $P(H|\overline{X})$ throughout analysis, to estimate both the significance of new results and the probability of obtaining similar results in future replications.

 \section{Significance and replication for the $t$ distribution}
 
 We consider an experiment involving  $N$ measurements which we assume follow a Normal distribution 
 \[X \sim \mathcal{N}(\mu,\sigma^2)\] 
 with unknown mean $\mu$ and unknown variance $\sigma^2$ (the standard assumption of normally distributed error), so that the sample mean $\overline{X}$ follows the distribution $ \mathcal{N}(\mu,\sigma^2/N)$.  Given  sample variance $S^2$ and degrees of freedom $\nu$, the variance in $\overline{X}$  is estimated by $S^2/N$ (the sample variance of the mean) and the standard approach  assesses significance against the point-form null hypothesis $\mu=0$ via a $t$ test.

  Letting $T_{\nu}$ represent the cumulative $t$ distribution with degrees of freedom $\nu$, $d=\overline{X}/S$  the normalised sample mean and $t=d\sqrt{N}$  the test statistic, in this approach the point-form significance of a result $t$ in a two-sided test is $p=2T_{\nu}(-|t|)$.    Since $d$ estimates the true normalised mean  $\mu/\sigma$ (a constant), the expected value of $|t|$ rises  with $N$ when $\mu \neq 0$, and so the expected value of $p$ falls with $N$ to a limit of $0$.     Since $\mu = 0$ can never hold exactly, in the standard approach to significance testing every experiment is expected to achieve significance for large enough $N$.   
 
 The sample variance of the mean in a given experiment is a function of the degree of random variation within that experiment.   This within-experiment variation arises as a consequence of  random factors affecting individual measurements in that experiment,  with each factor moving individual responses randomly one way or another.  If the number of such random confounding factors is large we can, via the central limit theorem, assume that the combined effect of all these factors will follow a Normal distribution (irrespective of the statistical distribution of the individual factors themselves),  so that the overall within-experiment variation is approximately Normal.  
 
 This within-experiment variation is, however, not the only form of variation in experimental means:  means are also affected by between-experiment variation (random variation in experimental means from one experiment to another).  This between-experiment variation  arises as a consequence of  factors that have a fixed or systematic effect within a given experiment, but vary randomly across different experiments with each factor moving the mean $\mu$ in a given experiment randomly one way or another.  Assuming that the number of such random between-experiment factors is also large, the overall between-experiment variation is, again, approximately Normal.   Between-experiment variation is very familiar to  researchers: a core aim in  experimental design is to control for  factors or confounds that may influence results. Such factors, however, can never be fully controlled (no experiment is perfect) and so necessarily cause some degree of random variation in experimental results: variation which should be taken into account when assessing significance.  
 
 Since the standard approach to significance testing only considers within-experiment variation (sample variance of the mean)  it systematically underestimates the degree of random variation in results, and so will identify results as `statistically significant' when in fact they arise as a consequence of random  between-experiment variation.  Our model assumes that $\mu$ itself varies randomly across experiments  following the distribution 
\[
 \label{eq:distributional_null} \mu \sim \mathcal{N}(\mu_0,\sigma_0^2)
\]
 with unknown variance $\sigma_0^2$, with the null hypothesis being $\mu_0=0$. 
 The variation of $\overline{X}$ around the overall mean $\mu_0$  now has two components: within-experiment variance $\sigma^2/N$ and between-experiment variance $\sigma_0^2$.  As $N$ increases the within-experiment variance of the mean, $\sigma^2/N$, falls and between-experiment variance  comes to be the dominant cause of random variation in means.   
   
 \subsection{Significance for a given variance ratio}
 We characterise the degree of between-experiment variance $\sigma_0^2$   via the variance ratio $b = \sigma_0^2/\sigma^2$.
 For a given value of $b$ we have 
 \[ 
 p(\overline{X}|N,\sigma, b) =  \int_{-\infty}^{\infty}\mathcal{N}(\overline{X}| \mu,\,\sigma^2/N) \mathcal{N}\left(\mu |\mu_0= 0,  b\sigma^2\right) \mathrm{d}\mu 
 =    \mathcal{N}\left(\overline{X}| 0,\, (\sigma^2/N)(1+ bN) \right)   
 \]
 and so
 \[
 \frac{\overline{X}}{(\sigma/\sqrt{N})\sqrt{1+bN}}
 \]
 is normally distributed with mean $0$ and variance $1$.  Letting $S^2$ be our sample variance (our estimate of $\sigma^2$) by definition
 \[  \frac{ S^{2}}{\sigma^{2}} \]
follows a $\chi_\nu^2$ distribution (where $\nu$ is the degrees of freedom $S^2$).  Letting
 \[ t = \frac{\overline{X}}{\sqrt{S^2} }\sqrt{N}\] 
as before and recalling that  the ratio of a standard Normal variable to the square root of a $\chi_\nu^2$ variable follows a $t$ distribution with $\nu$ degrees of freedom,     we   see that the statistic
 \begin{equation*}  
 \frac{ \frac{\overline{X}}{(\sigma/\sqrt{N})\sqrt{1+bN}}}{\sqrt{\frac{ S^{2}}{\sigma^{2}}}}  =\frac{t}{\sqrt{1+bN}} 
 \end{equation*}
  is $t$ distributed.  This means that the statistical significance of a given result $t$ in a two-sided test relative to the variance ratio $b$ is
\begin{equation} 
\label{eq:p_sig}  
 p_{sig}(t|b)= 2T_{\nu}\left(\frac{-|t|}{\sqrt{1+bN}} \right)
 \end{equation}
 Letting $t_{crit} =  T^{-1}_\nu(1-\alpha/2)$ be the critical value for significance in a two-sided test, a result $t$ will thus be significant at level $\alpha$  when 
 \[t_{crit} < \frac{|t|}{\sqrt{1+bN}} \]  
or, substituting  effect size $d=t/\sqrt{N}$,  when
\[t_{crit} < \frac{|d|}{\sqrt{b+1/N}} \]  
Since the $t$ distribution approaches the Normal distribution with rising $N$ and the sample effect size $|d|$ similarly approaches the true effect size $|\mu/\sigma|$ for our experiment,  we see that sample results from an experiment where

\[ \frac{|\mu/\sigma|}{\sqrt{b}} < \Phi^{-1}\left(1-\alpha/2\right)  \, \, \, \rightarrow \, \, \, |\mu/\sigma_0| < \Phi^{-1}\left(1-\alpha/2\right)\]
are never expected to reach significance at level $\alpha$, irrespective of sample size $N$ and, unlike with a point-form null, a distributional null hypothesis is not `always false'.

 \subsection{Significance, $\alpha$ and type $1$ error rate}
  The $p$ (in a point-form analysis) or $p_{sig}$ (in a distributional analysis) value of a given experimental result tells us the probability of a result as or more extreme arising under our (point-form or distributional) null hypothesis.  Given this probability a researcher must decide whether to treat this result as significant.   In Fisher's test for significance this decision is made by comparing the obtained probability against some subjective criterion $\alpha$, such that a probability less than $\alpha$ is counted as significant.   This criterion $\alpha$ represents the type 1 error rate of the test; that is, the probability of incorrectly rejecting the null hypothesis when that hypothesis is true.   
  
  Note that that this link between $\alpha$ and error rate holds independently of the form of the hypothesis: a point-form test $p < \alpha$ will incorrectly reject the point-form null hypothesis at rate $\alpha$, while a distributional test $p_{sig} < \alpha$ will incorrectly reject the \textit{distributional} null hypothesis at rate $\alpha$.   This does not, however, mean that point-form and distributional tests at some level $\alpha$ are in any way equivalent.  On the contrary, the fact that the distributional approach takes between-experiment variation into account while the point-form approach does not means that many results for which point-form $p < \alpha$ (suggesting a significant result) will simultaneously have $p_{sig} > \alpha$ (the result is not significant). Indeed, the difference between point-form and distributional significance can be extremely large: in our analysis of experiments in the Many Labs $1$ dataset (below), we found that around $10\%$ of experimental results with point-form $p < 0.00001$ (highly significant relative to the point-form null)  had $p_{sig}  > 0.05$ (not significant relative to the distributional null).

\subsection{Significance and effect size}
The size of an effect in a $t$ test is typically estimated in units of sample standard deviation, giving an effect-size estimate of $d=\overline{X}/S$ (an estimate of the true effect $\delta= \mu/\sigma$). 
Since in the point-form approach every experiment is expected to achieve significance for large enough $N$, effect size is necessarily independent of significance in this approach: the point-form significance of a result tells us nothing about its effect size.  Effect size is thus considered separately from significance, with researchers proposing various criteria to distinguish between strong, weak, and negligible effects \citep{cohen1992power,fritz2012effect,kelley2012effect,cohen2013statistical} alongside various lower bounds on acceptable effect sizes in terms of, for example, experimental `crud' \citep{meehl1990summaries,meehl1990appraising,orben2020crud}.  

In the distributional approach, by contrast, significance and effect size are no longer independent.  Instead  we have 
\[   
p_{sig}(t|b)= 2T_{\nu}\left(-|t|/\sqrt{1+bN}\right) = 2T_{\nu}\left(-|d|/\sqrt{b+1/N}\right) \approx 2 T_{\nu}\left(-|d|/\sqrt{b}\right)
\]
 and significance and effect size are directly related: if $p_{sig}(t|b)$ is low then the observed effect $d$ is large relative to the variance ratio $b$, while if $p_{sig}$ is high then the effect $d$ is small.  
A common recommendation in  effect size analyses is to select effect size criteria based on distributions of effect sizes for comparable outcome measures \citep{lipsey2012translating}.   This is just what  distributional significance provides: given an estimate of the between-experiment variance ratio $b$ (equivalent to a distribution of effect sizes),  in the distributional approach a significance level $\alpha$ places a bound on detectable effects.

\subsection{Power, $\beta$ and type $2$ error rate} 
Where $\alpha$ by construction estimates the type 1 error rate of an hypothesis test (the probability of incorrectly rejecting the null hypothesis when that hypothesis is true), the related measure $\beta$ estimates the type 2 error rate of that test (the probability of failing to reject the null hypothesis when it is in fact false), with $1-\beta$ being the statistical power of the experiment (the probability of correctly rejecting the null hypothesis when it is false).   Estimates of $\beta$ are made assuming that some true effect  $\delta=\mu/\sigma > 0$ holds; here we compare expressions for $\beta$ in point-form and distributional null approaches.

To estimate $\beta$ for tests relative to a point-form null hypothesis, we let $T_{\nu}(x;\theta)$ represent the cumulative probability from $-\infty$ to $x$ of the non-central $t$ distribution with degrees of freedom $\nu$ and non-centrality parameter $\theta$, and assume that $X \sim \mathcal{N}\left(\delta\sigma, \sigma^2\right)$ for some effect size $\delta \neq 0$ (so the null is false).  In this situation the variable $t=(\overline{X}/S)\sqrt{N}$ follows a non-central $t$ distribution with  non-centrality parameter $ \delta \sqrt{N}$.   The  probability, in the point-form approach, of getting a non-significant result in a two-sided test with critical value $t_{crit}$, given that effect $\delta$,  is then
\[ \beta= T_{\nu}\left( t_{crit};|\delta|\sqrt{N} \right) - T_{\nu}\left( -t_{crit};|\delta|\sqrt{N} \right)   \]
 and for an experiment to detect an effect of size $\delta$ with significance $\alpha$ and power $1-\beta$ we must have a sample size $N$ such that
\[ T_{\nu}\left(  t_{crit};|\delta|\sqrt{N} \right) - T_{\nu}\left( -t_{crit};|\delta|\sqrt{N} \right)\leq \beta \]
 Since    
\[\lim_{N \rightarrow \infty} T_{\nu}\left(  t;|\delta|\sqrt{N} \right) = 0\]
for all $t$ (as the non-centrality parameter approaches infinity, the area under the non-central $t$ distribution less than any fixed value $ t$ necessarily approaches $0$) and since
\begin{equation*}
\begin{split}
  \lim_{N \rightarrow 0}\left[ T_{\nu}\left(  t_{crit};|\delta|\sqrt{N} \right) -  T_{\nu}\left( -t_{crit};|\delta|\sqrt{N} \right) \right]
 = T_{\nu}\left(  t_{crit} \right) -T_{\nu}\left( - t_{crit} \right) = 1-\alpha
\end{split}
\end{equation*}
 (the non-central distribution approaching the central distribution in this situation)
we see that in the point-form approach $\beta$ has an upper limit of $1-\alpha$ and falls with increasing $N$ towards a lower limit of $0$.  The power of a given experiment ($1-\beta$) thus has a minimum of $\alpha$ and rises to a limit of $1$ with increasing $N$ (for any effect $\delta > 0$ there is a corresponding sample size $N$ which can detect that effect with any required power).

In the distributional situation we can do a similar analysis in terms of the overall null hypothesis by assuming some true effect $\delta=\mu_0/\sigma_0 > 0$, giving  
\[ 
\overline{X} \sim \mathcal{N}\left( \delta\sigma_0, (\sigma^2/N)(1+ bN) \right)   
\] 
and so
\[
\frac{\overline{X}}{(\sigma/\sqrt{N})\sqrt{1+bN}}
\]
is normally distributed with mean $\delta\sigma_0$ and variance $1$, and defining $t$ as before we see that 
\begin{equation*}  
\frac{t}{\sqrt{1+bN}} =\frac{ \frac{\overline{X}-\delta\sigma_0}{(\sigma/\sqrt{N})\sqrt{1+bN}} + \frac{\delta\sigma_0}{(\sigma/\sqrt{N})\sqrt{1+bN}} }{\sqrt{\frac{\nu\ S^{2}}{\sigma^{2}}/\nu}}   
\end{equation*}
follows a non-central $t$ distribution with $\nu$ degrees of freedom and non-centrality parameter 
\[  \frac{\delta\sigma_0}{(\sigma/\sqrt{N})\sqrt{1+bN}}  =  \frac{\delta}{\sqrt{1+1/bN}} \]
For a two-sided test we thus have a type $2$ error rate of
\[\beta =  T_{\nu}\left(t_{crit};\frac{|\delta|}{\sqrt{1+1/bN}}\right) - T_{\nu}\left(-t_{crit};\frac{|\delta|}{\sqrt{1+1/bN}}\right) \]
Here the non-centrality parameter falls towards $0$  with declining $bN$ and so the error rate $\beta$ rises, approaching an upper limit of
$1-\alpha$ as before.   As $bN$ rises, $\beta$ falls to a lower limit of
\begin{equation*}
\begin{split}
 & \lim_{bN \rightarrow \infty}  \left[T_{\nu}\left(t_{crit};\frac{|\delta|}{\sqrt{1+1/bN}}\right) - T_{\nu}\left(-t_{crit};\frac{|\delta|}{\sqrt{1+1/bN}}\right) \right] = T_{\nu}\left(t_{crit};|\delta|\right)-T_{\nu}\left(-t_{crit};|\delta|\right) 
  \end{split}
  \end{equation*}
 We thus see that the power of an experiment rises with $bN$ from a minimum of $\alpha$ towards a maximum of 
\[ 1- T_{\nu}\left(t_{crit};|\delta|\right) + T_{\nu}\left(-t_{crit};|\delta|\right) \approx 1- T_{\nu}\left(t_{crit};|\delta|\right)  \]
 (since the last term is negligible) and no experiment can detect an effect $\delta$ with a power greater than this value, irrespective of its sample size $N$ or the variance ratio $b$.  Since $T_{\nu}\left(t_{crit};|\delta|\right) > 0.5$ when $ |\delta| < t_{crit}$ this means that statistical power above $0.5$ can only be achieved for effects $|\delta|>  t_{crit}$, and that power approaches $1$ only for very large effect size $|\delta|$.  Where in the point-form approach we can design an experiment to detect any effect size at any required level of statistical power (by selecting a large enough $N$) in the distributional approach this is not possible, because the power of an experiment to detect an effect is, in the limit, a function of between-experiment variance rather than sample size.

\subsection{Estimating the probability of replication}
 Because this  distributional approach explicitly represents the variation in results across experiments (via the distributional null hypothesis), it allows us to express the probability of obtaining a statistically significant result at level $\alpha$ in a replication of our original experiment. In this section we give estimates for the replication probability, $p_{rep}$, for a given result based first on a fixed value $b$, and then based on a sample estimate $\hat{b}$.  
  
For some value of $b$ we assume a replication experiment designed to measure the same $X$ as the original (but with some between-experiment variation),  with  sample mean for the replication experiment following the distribution $\overline{X}_r \sim \mathcal{N}(\mu_r,\,\sigma_r^2/N_r)$, where $\mu_r$ and $\sigma^2_r$   are the unknown mean and unknown variance of $X$ in the replication, and $N_r$  and $\nu_r$  the number of measurements and the degrees of freedom.  We take $S_r^2$ to represent the sample variance of $X$ in the replication, and relate within-experiment variance in the original and replication via the ratio $c = \sigma_r^2/\sigma^2$.  

Letting $t_{crit_r} = T_{\nu_r}^{-1}(1-\alpha/2)$ be the critical $t$ value at level $\alpha$ in this replication experiment and
\[
t_r = \frac{\overline{X}_r}{ \sqrt{S_r^2} } \sqrt{N_r}   
\]   
the experimental result,  a result $t_r$ will count as a replication of our original result $t$ if both $t$ and $t_r$ are significant at the same level $\alpha$, and in the same direction, relative to our distributional null hypothesis.
In other words, if our original result $t$ was significant in a left-tailed test ($t < -t_{crit}\sqrt{1+bN}$) then it is replicated if $t_r < -t_{crit_r}\sqrt{1+bN_r}$, while if our original result was significant in a right-tailed test ($t > t_{crit}\sqrt{1+bN}$) then it is replicated if $t_r > t_{crit_r}\sqrt{1+bN_r}$.  
 
To estimate the probability of replication we must derive a distribution for $t_r$.  We use Bayes' theorem to update our initial null hypothesis distribution $\mu \sim \mathcal{N}( 0,\,\sigma_0^2)$ using  the observed value of the sample mean in the original experiment, $\overline{X}$, getting the posterior distribution for $\mu$ of
 \[
 \mu \sim \mathcal{N}(  \mu_N,\,\sigma_N^2)  
 \]
 where
 \[ 
 \sigma_N^2 = \frac{\sigma_0^2\,  \sigma^2 }{N\, \sigma_0^2 + \sigma^2}  =  \frac{bN}{(1+bN)}  ( \sigma^2/N)   
 \]
 \[ 
 \mu_N =  \frac{\sigma_N^2  }{( \sigma^2/N)}\, \overline{X}=  \frac{bN}{(1+bN)} \overline{X}
 \]
    \citep[a property of the Normal distribution; see e.g.][]{bishop2006pattern,murphy2007}.  Note that this distribution  does
    not represent an unconditional inference about the probability distribution for $\mu$: our experimental result does not tell us that $\mu$ in fact has this distribution (because a distribution for $\mu$ cannot be inferred from a single mean).  Instead this is the inferred distribution for $\mu$ conditional on the initial null hypothesis; that is, conditional on the initial assumption that $\mu \sim \mathcal{N}( 0,\,\sigma_0^2)$.   
    
    Given this updated distribution for $\mu$, the probability density of the sample mean $\overline{X}_r$ in our second experiment is
 \[
 p(\overline{X}_r|\overline{X},  b) =  \int \mathcal{N}(\overline{X}_r| \mu,\,c \sigma^2/N_r) \mathcal{N}(\mu | \mu_N,\,\sigma_N^2) \mathrm{d}\mu
 \]
 and so 
 \[
 \overline{X}_r  \sim \mathcal{N}( \mu_N,\,c\sigma^2/N_r+\sigma_N^2) 
 \sim \mathcal{N}\left( \frac{bN}{(1+bN)}\, \overline{X},\,   ( \sigma^2/N) \left(    \frac{bN}{1+bN}+\frac{cN}{N_r}\right) \right) 
 \]
 and the variable
 \[
 \frac{\overline{X}_r-\frac{bN}{1+bN}\, \overline{X}}{\frac{\sigma}{\sqrt{N}}\sqrt{  \frac{bN}{1+bN}+\frac{c N}{N_r}  }} 
 \]
 is normally distributed with mean $0$ and variance $1$.  This means that the variable 
 \[ 
 \label{eq:rep_t} 
 \frac{\overline{X}_r-\frac{bN}{1+bN}\, \overline{X}}{\sqrt{(\frac {\nu_r\ S_r^{2}}{c \sigma^{2}})/\nu_r}  \left(\frac{\sigma}{\sqrt{N}}\sqrt{ \frac{bN}{1+bN}+\frac{c N}{N_r}}\right)} 
 =  \frac{t_r\sqrt{ \frac{c N}{N_r}  }-\frac{bN}{1+bN} \frac{\overline{X}}{\sqrt{S^2_r/N}}\sqrt{c}}{\sqrt{  \frac{bN}{1+bN}+\frac{c N}{N_r}  }}
 \]  
 has a $t$ distribution with $\nu_r$ degrees of freedom.

 Since the variance in the replication experiment is  $c$ times that of the original experiment, the variable $S^2_r/c$ follows approximately the same distribution as the variable $S^2$ (a scaled $\chi^2$ distribution) when  $N$ and $N_r$ are similar (with both variables following exactly the same distribution when $N=N_r$), or when both are large.  In this situation the variable
 \[   
 \frac{t_r\sqrt{ \frac{c N}{N_r}  }-\frac{bN}{1+bN} \frac{\overline{x}_1}{\sqrt{S/N}}}{\sqrt{  \frac{bN}{1+bN}+\frac{c N}{N_r}  }} =   \frac{t_r\sqrt{ \frac{c N}{N_r}  }-\frac{bN}{1+bN}t}{\sqrt{  \frac{bN}{1+bN}+\frac{c N}{N_r}  }} 
 =   \frac{t_r  \sqrt{c}-t\, \frac{b \sqrt{N\, N_r}}{1+bN}}{\sqrt{  c+bN_r/(1+bN)}} 
 \]
 will follow a $t$ distribution with $\nu_r$ degrees of freedom.

 Given this distribution the probability of obtaining a result less than $-t_{crit_r}\sqrt{1+bN_r}$ in our replication experiment is 
\begin{equation}
\label{eq:crit_{rep}_left}
  T_{\nu_r}\left(  \frac{-t_{crit_r}  \sqrt{c(1+bN_r)}-t\, \frac{b \sqrt{N\, N_r}}{1+bN}}{\sqrt{  c+bN_r/(1+bN)}} \right) 
\end{equation}  
 and the probability of obtaining a result greater than $t_{crit_r}\sqrt{1+bN_r}$ is
 \begin{equation}
 \label{eq:crit_{rep}_right}
1- T_{\nu_r}\left(  \frac{t_{crit_r}  \sqrt{c(1+bN_r)}-t\, \frac{b \sqrt{N\, N_r}}{1+bN}}{\sqrt{  c+bN_r/(1+bN)}} \right) = T_{\nu_r}\left(  \frac{t\, \frac{b \sqrt{N\, N_r}}{1+bN} - t_{crit_r}  \sqrt{c(1+bN_r)}}{\sqrt{  c+bN_r/(1+bN)}} \right) 
 \end{equation}  
 If our original result $t$ was negative, then the probability of replication is given by Equation \ref{eq:crit_{rep}_left} while if positive it is given by Equation  \ref{eq:crit_{rep}_right} and so substituting for $t_{crit_r}$ we see that the probability of replication of any result $t$ (given variance ratios $b$ and $c$) is
\begin{equation*}
 T_{\nu_r}\left(  \frac{|t|\frac{ b \sqrt{N\, N_r}}{1+bN} - T^{-1}_{\nu_r}\left(1-\alpha/2\right)\sqrt{c(1+bN_r)}  }{ \sqrt{  c+bN_r/(1+bN)  }  } \right) 
\end{equation*} 
Since our second experiment is a replication of the first, we can generally assume that the within-experiment variance $\sigma^2$ in both experiments will be approximately the same (so $c=1$) and, simplifying, we have an expression for the probability of replication of
\begin{equation}
\label{eq:crit_{rep}}
p_{rep}(t|b) = T_{\nu_r}\left(  \frac{|t| b \sqrt{N\, N_r} - T^{-1}_{\nu_r}\left(1-\alpha/2\right)(1+bN)\sqrt{1+bN_r}  }{ \sqrt{ (1+bN+bN_r)(1+bN)  }  } \right) 
\end{equation} 

\subsection{Bounds on significance and replication}
These expressions impose useful bounds on significance and replication probabilities for a range of between-experiment variance values $b$. 
From Equation \ref{eq:p_sig} we see that the distributional significance of a given result $p_{sig}(t|b)$ falls with falling $b$, so that if a result is significant at level $\alpha$ for some variance ratio $B$, it is significant at the same level for all $b < B$.  A similar result holds for $p_{rep}(t|b)$: it can be shown that the probability of replication of a given result $p_{rep}(t|b)$ \textit{rises} with falling $b$ to a maximum at some value $b_{max} \approx |d|/\sqrt{N}$, so that $p_{rep}(t|b) \geq p_{rep}(t\mid B)$ for almost all $b \leq B$ (see Appendix $1$ for proof).  If we have some reasonable upper limit $B$ on the variance ratio for experiments (such that the actual unknown variance ratio for any given experiment is \textit{a priori} expected to be less than this maximum), then we can define `generic' estimates for distributional significance and replication as 
\[
\hat{p}_{sig} = 2T_{\nu}\left(-|t|/\sqrt{1+BN}\right)
\]  
and
\[
\hat{p}_{rep}=T_{\nu_r}\left(  \frac{|t| B \sqrt{N\, N_r} - T^{-1}_{\nu_r}\left(1-\alpha/2\right)(1+bN)\sqrt{1+BN_r}  }{ \sqrt{ (1+BN+BN_r)(1+BN)  }  } \right)
\] 
 and if for a given result $t$ we have $\hat{p}_{sig} \leq \alpha$ and $\hat{p}_{rep} = \rho$, then our result $t$ is significant at level $\alpha$, and has a probability of replication greater than $\rho$, conditional on the assumption that  between-experiment variance is less than $B$ times within-experiment variance.

What value should be chosen for this bound $B$?    Differences in means across experiments can arise as a consequence of  within experiment variance and as a consequence of differential factors affecting experimental means (if the means in one experiment is affected by some confounding factor not present in the other, we expect their means to differ systematically as a function of that confound).    Since replication experiments are necessarily designed to be as similar as possible to each other, we expect the effect of these between-experiment confounds  to be weaker than the effect of within-experiment variance (if between-experiment confounds have a strong influence on experimental means, then the experiments are not, in fact, replications: they differ in some important aspect).   Given that the variance ratio $b$ represents the ratio of between-experiment to within-experiment variance, we thus expect that   $b < 1$ will hold for replications and set a maximum bound of $B=1$. 

Note, however, that the lower the value of $B$, the lower the value of $\hat{p}_{sig}$ and the higher the value of $\hat{p}_{rep}$ (and so the more likely we are to judge a result as significant and replicable).  This means that lower values of $B$ (those that may underestimate the true level of between-experiment variance) have a higher type $1$ error rate but a lower type $2$ rate, while higher values of $B$ (that may overestimate the true level of between-experiment variance) have a lower type $1$ error rate and a higher type $2$ rate.  For conservative rejection of the null (only rejecting the null when the evidence is very strong), we should therefore test against a bound of $B$ close to $1$; for more liberal rejection of the null (being willing to reject the null even when the evidence is weak), we should test against a bound of $B$ close to $0$.

\section{Estimating significance and replication}

The previous section derived distributional null expressions for significance, power and replication assuming known variance ratio $b$ or bounds on that ratio.  These expressions are both computationally complex, and depend on this unknown value.  In this section we derive sample estimators for $b$, and give simple and easily used closed-form estimates for significance and replication in terms of the standard $t$ distribution.  We also describe extensions of this approach to different forms of $t$ test and to tests of regression, linear correlation, and contingency.

\subsection{Sample Estimators for $b$}
Our estimates $\hat{p}_{sig}$ and $\hat{p}_{rep}$ are in terms of generic bounds on $b$. Given replication data, however, we can estimate $b$ in terms of its component variances, producing more specific expressions for $p_{sig}$ and $p_{rep}$. 
Estimating $\sigma^2$ is straightforward: given some experiment measuring a normally distributed variable $X \sim \mathcal{N}(\mu_i,\,\sigma_i^2)$ with unknown within-experiment variance $\sigma_i^2$ and degrees of freedom $\nu$, the unbiased estimator for $\sigma^2$ is  
\[	S_i^2 = \frac{1}{\nu}\sum_{i=1}^N(X_i-\overline{X})^2 
\]
the sample variance of $X$. We wish to give an unbiased estimator for $\sigma_0^2$, which we write as  $S_0^2$.  We obtain this estimator by considering the variance  of the means in some set of replications of an experiment similar to ours, and subject to the same broad set of random confounding factors, both within and between experiments.  Since in psychology many such factors are related to differences between individual participants (for within experiment variance) and to systematic differences between groups of participants (for between experiment variance), such estimates of the between-experiment variance of means can potentially be applied  widely.

Suppose we have $K$ experiments  all measuring the same variable $X$.  In each experiment $i$ we assume that $X  \sim \mathcal{N}(\mu_i,\,\sigma_i^2)$ for different values of $\mu_i$ and $\sigma_i^2$, where $\mu_i \sim \mathcal{N}(\mu_0,\,\sigma_0^2)$ ($\sigma_0^2$ representing the between-experiment variance of the mean).  We assume that each experiment $i$ has $N_i$ measurements of $X$ with sample mean $\overline{X}_i$  and within-experiment sample variance $S_i^2$, and take $\overline{X}_*$ to the mean of these sample means $\overline{X}_i$ and
\[  S^2_* = \frac{1}{K-1}\sum_{i=1}^K(\overline{X}_i-\overline{X}_*)^2\]
to be the sample variance of these sample means.  Then since each $\overline{X}_i$ is distributed as 
\[ \overline{X}_i \sim \mathcal{N}(\mu_i,\sigma_0^2+\sigma_i^2/N_i)\]
the theoretical variance of their overall mean $\overline{X}_*$ is given by
\[   \frac{1}{K} \sum_{i=1}^K (\sigma_0^2+\sigma_i^2/N_i)  = \sigma_0^2 + \frac{1}{K} \sum_{i=1}^K \sigma_i^2/N_i\]
and equating variances and noting that $S^2_i$ is an unbiased estimator of $\sigma^2_i$ this gives an estimate for $\sigma_0^2$ of
\[ S_0^2 = S^2_* -  \frac{1}{K} \sum_{i=1}^K S_i^2/N_i\] 
This is the Hedges estimator for heterogeneity in meta-analysis, and is an unbiased estimator of $\sigma_0^2$, even when $\sigma_i^2$ is unknown and is estimated by $S_i^2$ \citep[see e.g.][]{hedges1983random,viechtbauer2005bias,veroniki2016methods}.  Since the within-experiment sample variance of the mean $S_i^2/N_i$ declines to $0$ with increasing  $N_i$, the expression $(K-1)S_0^2/\sigma_0^2$ approximately follows a $\chi^2$ distribution with degrees of freedom $\nu_0=K-1$ (with the error of the approximation falling with increasing $N_i$).  

Note that in the case where we have only seen $1$ experiment with mean $\overline{X}$ and sample variance $S^2$ and no data on the variation of sample means across experiments, the above expression for $S_*^2$ is undefined and so an estimate for $S_0^2$ cannot be given.  Similarly, since estimates for $\sigma_0^2$ necessarily depend on accurate estimates of the component variances $\sigma_i^2$, when  $S^2_0 < 0$ we can conclude that some sample sizes $N_i$ are too small to give the required accuracy; and again, an estimate for $S_0^2$ cannot be given.  

Given unbiased estimators $S^2$  and $S_0^2$ we take $ \hat{b} = S_0^2/S^2$ as our estimate for the ratio $b=\sigma^2_0/\sigma^2$, and because both $S^2$  and $S_0^2$ approximately follow $\chi^2$ distributions, we see that $b$ approximately follows an $F_{\nu,\nu_0}$ distribution scaled by this estimate $\hat{b}$ (with the error in the approximation falling with $\nu_0$).  The variable $p_{sig}$ is then distributed as a function of the distribution of $b$, and the expected value of that distribution given $ \hat{b}$ is   
\[
p_{sig}(t|\hat{b}) = \int_{0}^{\infty} 2T_{\nu}\left(\frac{-|t|}{\sqrt{1+b \hat{b} N}}\right) f_{\nu,\nu_0}(b) \mathrm{d}b 
\] 
where $f_{\nu,\nu_0}$ is the probability density of that $F$ distribution.  
This is our theoretical model for distributional significance, and when $p_{sig}(t|\hat{b}) < \alpha$ we can say that our result $t$ gives statistically significant evidence against the null hypothesis, given the estimated within-experiment and between-experiment variances.  

To consider the probability of replication in a new experiment, we assume that sample variances $S^2$ and $S_r^2$ in the two experiments are both estimates of the same underlying within-experiment variance $\sigma^2$ as before.  Then since these two sample variances  both follow a $\chi^2$ distribution with the same mean and with degrees of freedom $\nu$ and $\nu_r$,  their ratio $c$ follows an $F$ distribution $F_{\nu,\nu_r}$.  Similarly taking $ \hat{b} = S_0^2/S^2$ as our estimate for the ratio $b$ we see that the variable $p_{rep}$  is  distributed as a function of the joint distribution of $b$ and $c$. Since we don't know the sample variance $S_r^2$ (our replication experiment has not yet been carried out) we simply set $\hat{c}=1$ and the expected value for $p_{rep}$ given estimate $\hat{b}$ is
\begin{equation}
\label{eq:p_rep_hat_b}
p_{rep}(t|\hat{b}) =    \int_{0}^{\infty} \int_{0}^{\infty} T_{\nu_r}\left(  \frac{|t|  \sqrt{N\, N_r} - T^{-1}_{\nu_r}\left(1-\alpha/2\right)(1+b \hat{b}N)\sqrt{1+b \hat{b}N_r}  }{ \sqrt{ (1+b \hat{b}N+b \hat{b} N_r)(1+b \hat{b}N)  }  } \right)
  f_{\nu,\nu_0}(b) f_{\nu,\nu_{r}}(c)\,  \mathrm{d}b\, \mathrm{d}c
\end{equation}
This is our theoretical model of the probability of result $t$ being replicated in an experiment with $N_r$ measurements, given an estimate of between-experiment variance in the means of a set of similar experiments and assuming the sample variance in that replication is close to $S^2$.   Note that this estimate for replication is conditional on evidence against that null hypothesis from the experimental result $t$.    This means that if result  $t$  does not give evidence against the null hypothesis (that is, if $p_{sig}$ is high) then we are not justified in `updating away' from that hypothesis, and our estimate of $p_{rep}$ does not apply. In other words, we only have evidence for replication when we have evidence against the null. 

These integral and double-integral estimates for $p_{sig}$ and $p_{rep}$ are, however, somewhat computationally expensive to calculate in practice; and further, these calculations can be `fragile': dependent in opaque ways on the numerical integration algorithm used, especially for values of $\hat{b}$ close to $0$.  In the next section we give more practically useful estimates for $p_{sig}$ and $p_{rep}$, expressed in terms of the standard $t$ distribution.

\subsection{Closed-form significance and replication estimates}
The difference between point-form and distributional significance testing is most important for large sample size $N$, because in that situation between-experiment variance $\sigma_0^2$ dominates the sample variance of the mean $\sigma^2/N$.  Here we derive approximate estimates for significance and replication for large values of $N$: specifically, values such that 
$S^2$ is approximately equal to $\sigma^2 $ and such that $\sigma/\sqrt{N}$ is much smaller than $\sigma_0$, which means that
\[ \frac{bN}{1+bN} \approx 1\]
Then we have
\[
p_{sig}(t|b) =2T_{\nu}\left(\frac{-|t|}{\sqrt{1+b N}}\right) \approx 2\Phi\left(\frac{-|t|}{\sqrt{1+b N}}\right) \approx 2\Phi\left(\frac{-|\overline{X}|} {(\sigma/\sqrt{N})\sqrt{\frac{\sigma_0^2}{\sigma^2} N} } \right) = 2\Phi\left(\frac{-|\overline{X}|}{\sqrt{\sigma_0^2} }\right)  
\] 
Since by assumption $\sigma/\sqrt{N}$ is much smaller than $\sigma_0$, we see that $\overline{X}$ approximately follows the distribution $\mathcal{N}(0,\sigma_0^2)$.  Since  $S_0^2$ is an unbiased sample estimator for $\sigma_0^2$   such that $\nu_0S_0^2/\sigma_0^2$ approximately follows the distribution $ \chi^2_{\nu_0}$, this means that
\[ \Phi\left(\frac{-|\overline{X}|}{\sqrt{\sigma_0^2} }\right) \approx T_{\nu_0}\left(\frac{-|\overline{X}|}{\sqrt{S_0^2} }\right) =   T_{\nu_0}\left(\frac{t}{\sqrt{\hat{b}N}}\right) \]
(Note the change in degrees of freedom for the $t$ distribution, from $\nu$ to $\nu_0$), and we define our general expression for distributional significance as
	\[
	p_{sig} = 2T_{\nu_0}\left(\frac{-|t|}{\sqrt{\hat{b}N}}\right)
	\]
For replication we have  
\begin{equation*}
\begin{split}
p_{rep}(t|b) &= T_{\nu_r}\left(  \frac{|t|b \sqrt{N\, N_r}}{ \sqrt{  (1+bN)(1+b(N+N_r))  }  } - \frac{T^{-1}_{\nu_r}\left(1-\alpha/2\right)\sqrt{(1+bN)(1+bN_r)}  }{ \sqrt{1+b(N+N_r)  }  } \right) \\
\end{split}
\end{equation*}  
and since  by assumption $\sqrt{1+bN} \approx \sqrt{bN}$, we have the approximation
\begin{equation*}
\begin{split}
p_{rep}(t|b) &\approx T_{\nu_r}\left( \sqrt{\frac{N_r}{N+N_r}}\left[  |t|  - T^{-1}_{\nu_r}\left(1-\alpha/2\right)\sqrt{1+bN} \right]\right) \\
\end{split}
\end{equation*} 
By assumption we also have $S^2 \approx \sigma^2$ so that 
$b = \sigma_0^2/\sigma^2 \approx \sigma_0^2/S^2$
and so we can rearrange to get
\[ p_{rep}(t|b) \approx T_{\nu_r}\left(\sqrt{\frac{\hat{b} N N_r}{N+N_r  } }\left[ \frac{|t|}{\sqrt{\hat{b}N}} -
T^{-1}_{\nu_r}\left(1-\alpha/2\right)\sqrt{ \frac{1}{\hat{b}N}  + \frac{\sigma_0^2}{S_0^2}}   \right] \right) 
\]
Since  $\sigma_0^2/(\nu_0\,S_0^2)$ approximately follows the inverse distribution $\chi^{-2}_{\nu_0}$, and this distribution has a mean of $1/(\nu_0-2)$ for $\nu_0 > 2$, this gives an expected replication probability of
\[ p_{rep}  =  T_{\nu_r}\left(\sqrt{\frac{\hat{b} N N_r}{N+N_r  } }\left[ \frac{|t|}{\sqrt{\hat{b}N}} -
T^{-1}_{\nu_r}\left(1-\alpha/2\right)\sqrt{ \frac{1}{\hat{b}N}  + \frac{\nu_0}{\nu_0-2}}   \right] \right)
\]
which defines our general expression for replication.

These $p_{sig}$ and $p_{rep}$ expressions are closed-form estimates for distributional significance and replication, analogous to the point-form $t$ test for significance in that they use the $\chi^{2}$ distribution to replace the unknown variance $\sigma^2_0$ with its estimate $S_0^2$.  We expect these expressions to be less accurate in terms of estimating significance and replication probabilities than the full integral expression given in Equation \ref{eq:p_rep_hat_b} (with the difference falling as sample size $N$ rises) but these closed-form estimates have the advantages of simplicity,  computational speed, and robustness.    

These expressions also clarify the general relationship between distributional significance and replication: given a result $t$ which exactly meets the criterion for significance we have  
\[|t| = T^{-1}_{\nu_0}\left(1-\alpha/2\right)\sqrt{\hat{b}N}\]
and so substituting the probability of replication for this result is
\[ p_{rep}(t|b) =  T_{\nu_r}\left(  T^{-1}_{\nu_0}\left(1-\alpha/2\right)  \, \sqrt{\frac{\hat{b} N N_r}{N+N_r  } }\left[ 1-
\sqrt{ \frac{1}{\hat{b}N}  + \frac{\nu_0}{\nu_0-2}}   \right] \right)
\]
The innermost term is negative and approaches $0$ with rising  $\nu_0$ and $\hat{b}N$, and so $p_{rep}$ for such a result is less than, but approaching, $0.5$.  This means that non-significant results necessarily have $p_{rep} < 0.5$ in this model and significant results have $p_{rep} \geq 0.5$ (except in limiting cases where $N$ is small and $p_{sig}$ is close to $\alpha$).  In other words, where the standard point-form significance measure $p$ tells us nothing about the probability of replication of a given result, the distributional significance measure $p_{sig}$ gives direct information about the probability of replication: if $p_{sig} < \alpha$ for a given result then we expect that result to replicate at least $50\%$ of the time (with the probability of replication increasing as $p_{sig}$ falls).

 \subsection{Extension to related statistical tests}
  We've presented distributional significance testing here in terms of a generalised $t$ test, and so this approach applies to each distinct form of that test. To apply this model to a one-sample $t$ test on a variable $X$ with sample size $N$, we set $\nu=N-1$ and take $S^2$ to be the sample variance of that variable.  To apply to a paired $t$ test on two variables (assuming $N$ pairs) we take the $X$ to represent the difference in pairs and $\overline{X}$ and $S^2$ to be the average difference between pairs and the variance of that difference, and $\nu=N-1$.   To apply to an unpaired or independent $t$ test on two variables (with $N$ measurements in total), we take   $\overline{X}$ to be the difference between the means of each variable, $N$ to be the sample size, and  $S^2$ to be the pooled sample variance with degrees of freedom $\nu=N-2$. (Note that we implicitly assume an equal number of samples for the two variables in an unpaired test.  This is because the probability of replication of a given result in a future experiment on two unpaired variables will depend on the sample sizes for those two variables in that experiment, which we cannot know.  Assuming that these sample sizes vary randomly the expected sample size for each variable will be $N/2$; and so we use the calculated replication probability assuming equal sample sizes as our estimate of the probability of replication.)    
  
  The approach also applies to tests of linear regression, correlation and contingency, which can all be related to the $t$ test.  Applying the distributional approach to tests of independence for correlation and linear regression gives $p_{sig}$ and $p_{rep}$ expressions as above, albeit calculated with sample slope $M$ replacing sample mean $\overline{X}$, predictor sum-of-squares (which we write as $Q$) replacing sample size $N$, and mean squared error of the response estimate replacing sample variance. This is because, with these replacements, such tests of independence are simply $t$ tests (see Appendix $2$ for a derivation of these results).  Note that in these correlation tests the sum of squares of the predictor variable is the continuous-valued analog of the discrete sample size  $N$ in a $t$ test, and is treated as a controlled variable: where sample size gives the number of discrete measurements made  in a $t$ test,  predictor sum-of-squares gives the range of continuous measurements  in a test of correlation.   This means that the significance and replication probability of these tests depends on the choice of predictor and response variables: an estimate for the probability of replication of a given result in an experiment where $X$ is the predictor (and so assumed to be controlled in replications) will necessarily be different from the probability of replication of that result when $Y$ is the predictor and the $Y$ sum of squares is assumed to be controlled in replications (because the two replication experiments themselves are different).   
  
 Code to calculate all these tests of distributional significance and replication is available online at \url{https://osf.io/ep86y}.   This code also tests the distributional model's approach to significance and replication against the Many Labs 1 replication dataset, as described in the next section.

  \subsection{Comparing with point-form approaches to replication} 
  How does this distributional model compare with other approaches to replication? The most influential method for estimating the probability of replication $p_{rep}$ is that of Killeen  \citep{killeen2005alternative,killeen2005replicability,killeen2007replication,sanabria2007better}, which for a number of years was recommended by
  the journal \textit{Psychological Science} as an alternative to standard hypothesis tests.   This approach does not estimate the probability of obtaining a statistically significant result in a second experiment: instead it estimates the probability of obtaining a result in a second experiment that is  the same sign as the original result.  In a one-sample test with sample size $N$ and effect $d=\overline{X}/S$, Killeen's $p_{rep}$ calculates the probability of obtaining a result of the same sign as $d$ by estimating the variance of the original effect as $\sigma^2_d \sim N/(N-4)$, taking the variance in the ensemble of original and replication to be $2\sigma^2_d$, and assuming that effects are normally distributed, so that Killeen's $p_{rep}$ for a given $d$ is approximately
  \begin{equation*}
  \Phi\left(\frac{|d|}{\sqrt{2}\sigma_d}\right) = \Phi\left(\frac{|\overline{X}|}{\sqrt{2}S} \left[1-\frac{4}{N}\right]\right)
  \end{equation*}

  Researchers have pointed out a number of fundamental problems with this approach to replication: that there are errors in the derivation of Killeen's $p_{rep}$  \citep{doros2005probability}; that  Killeen's $p_{rep}$ is simply a function of the point-form $p$ and so gives no further information \citep{maraun2010killeen};  that  Killeen's $p_{rep}$ estimates an aggregated probability over all results and experiments, rather than the probability of a given result being replicated in a particular experiment \citep{miller2009probability,miller2011aggregate}; that replication probabilities must be estimated based on a prior distribution \citep{macdonald2005replication}; that Killeen's $p_{rep}$ assumes all effect sizes are equally likely and so assumes \textit{a priori} that the null hypothesis is false \citep{iverson2010model};  and finally that the claims associated with Killeen's $p_{rep}$ are  easily misinterpreted  \citep{iverson2009prep,iverson2009fits,iverson2010random}.
  
  Our distributional null hypothesis expression for $p_{rep}$ resolves the problems in Killeen's approach.  Our $p_{rep}$ is not simply a function of the point-form $p$ and is  an estimate for replication in a particular experiment (in that it depends on the  ratio of between-experiment to within experiment variance $b$, the ratio of within-experiment variance in the original experiment and the replication $c$, and the  sample sizes of the original and replication experiments).  Our $p_{rep}$ is specifically based on a distribution for the null hypothesis, does not assume all effect sizes are equally likely (the larger an effect size the less likely it is under this distribution), and does not assume the null hypothesis is false.  Finally, since our result is based on well-known properties of the Normal and $t$ distributions, errors in the derivation are (hopefully) avoided, and the associated claims have a clear interpretation.

\section{ Predicting Replication in the Many Labs 1 Dataset}
We test this distributional  approach  using data from the first `Many Labs' replication project \citep{klein2014investigating}.  This  involved the replication of $16$ different experiments investigating a variety of classic and contemporary psychological effects.  Each experiment was originally published in the cognitive or social psychology literature, and was replicated by researchers in around $36$ different sites ($25$ in the US, $11$ international): each individual instance of one of these tasks thus had around $35$ replications with the same design but with varying sample size and sample variance values (some tasks had less than $36$ experiments; there were $574$ experiments in total, with $400$ occurring at US sites).   We chose to use this dataset to test the distributional null approach because the raw response data for all experiments at all sites is available, allowing us to compute the various estimators required in the distributional model, and because the large number of replications of each individual task allows an accurate estimate of the actual replication rate for these experiments, which we can use to test the replication probability predicted for those experiments by our model.     Note that this dataset does not include published results from any of the original experiments being tested, but only results from planned replications across the $36$ sites (so minimising the impact of publication bias).

These experiments covered a wide range of topics, including the impact of question phrasing on whether speeches for or against democracy should be allowed/forbidden;  the effect of `anchoring' on estimates of the distance from San Francisco to NYC, the number of babies born per year in the US, the population of Chicago, or the height of Mount Everest;   agreement with quotes depending on they are attributed to George Washington/Osama Bin Laden; the effect of priming with the US flag/currency on political opinions; and the effect of imagined contact with Muslims on reductions in prejudice.  A number of these topics were US-specific, and so experimental ``replications'' of these results in non-US sites are subject to clear confounds (we would expect there to be differences between US and non-US sites in knowledge about US distances and US city populations, and in attitudes towards George Washington and the US flag or currency, for example\footnote{Such differences are evident even in knowledge about the height of Mount Everest: on average fewer than $2\%$ of estimates at US sites fell within $1\%$ of the true height, with all US sites having less than $6\%$ of estimates in that range.  Of the $11$ international sites, however, one had $35\%$ of estimates  within $1\%$ of the true height, one $26\%$ and one $20\%$, and on average more than $10\%$ of international estimates were within $1\%$ of the true height.}).  For this reason we limit our analysis to results from US sites only.

 \subsection{Methods}   
 All data, analysis and visualisation code is publicly available on the OSF repository \citep{Costello_2022}. Data were analyzed using R, version $4.0.5$ \citep{Rmanual2021} and the packages ggplot2 version $3.3.5$ \citep{ggplot2_2016} and xtable $1.8.4$ \citep{dahl2009xtable}.  The R analysis script downloads the Many Labs $1$ dataset  from the project's OSF repository and carries out all analyses described below.  
 
 Experiments in the Many Labs dataset varied widely in significance, with some having significant results in all or almost all replications, and others having only a few significant results.     In our analysis we use the independent and dependent measures from the Many Labs 1 tasks (and the corresponding criteria for inclusion/exclusion of responses as flagged in the dataset);   Table \ref{tab:replication} gives summary information on these tasks for US sites.

 \begin{table}[t!]
	\caption{\label{tab:replication}  Summary information for US sites in Many Labs 1 replication project, with $25$ individual experiments for each experimental task. The table also gives the estimated sample variance of the means across experiments in each task ($S^2_0$), the proportion of experiments in each task that were statistically significant  in two-sided point-form  and distributional tests ($p$ or $p_{sig}$ less than $0.05$) and where the effect was in the same direction as in the original experiment (this is equivalent to significance at $0.025$ in a one-sided point-form or distributional test  with the direction of the test corresponding to that of the original effect).  The $p$ and $p_{sig}$ values of the aggregated dataset for each experimental task are also shown (with between-experiment variance for a given task taken to be $S^2_0$ and within-experiment variance calculated from the aggregated dataset for that task).     }
	\centering 
	\scalebox{0.8}{
		\begin{tabular}{rp{5.3cm}c @{\extracolsep{0.05cm}}c c cccc @{\extracolsep{0.0cm}}r c @{\extracolsep{0.0cm}}r c}
			\hline	 
			&   &    & &  & &   \multicolumn{4}{c}{significant at  $\alpha=0.05$ and in}   \\  
			& Task  & Total   & test & $S_0^2$ & $\hat{b}$    &\multicolumn{4}{c}{same direction as original} \\
			\cline{7-10}
			&   & participants   & &  &   &\multicolumn{2}{c}{proportion} &\multicolumn{2}{c}{\ \ \ aggregated \ \ \ }   \\ 
			\cline{7-8} 	\cline{9-10}
			&	&   &  &  &(mean) & $p$ & $p_{sig} $ &  $p$ & $p_{sig}$   \\ 
			\hline \\
 1 & Allowed/forbidden & 4934 & $\chi^2$ & 0.005 & 0.041 &   1.00 &   1.00 & $10^{-325}$ & $10^{-10}$ \\ 
2 & Anchoring - babies born & 4481 & $t$ & $6.4 \times 10^{6}$ & 0.06 &   1.00 &   1.00 & $10^{-325}$ & $10^{-4}$ \\ 
3 & Anchoring - Mt. Everest & 4575 & $t$ & $9.3 \times 10^{6}$ & 0.11 &   1.00 &   1.00 & $10^{-325}$ & 0.0004 \\ 
4 & Anchoring - Chicago & 4302 & $t$ & $8.1 \times 10^{10}$ & 0.077 &   1.00 &   1.00 & $10^{-325}$ & 0.001 \\ 
5 & Anchoring - distance to NYC & 4422 & $t$ & $5.0 \times 10^{4}$ & 0.05 &   1.00 &   0.76 & $10^{-266}$ & 0.01 \\ 
6 & Relation between Implicit and Explicit math attitudes & 4383 & $r$ & 0.080 & 0.12 & 0.88 & 0.52 & $10^{-148}$ & 0.01 \\ 
7 & Retrospective gambler fallacy & 4690 & $t$ & 0.640 & 0.12 & 0.88 & 0.00 & $10^{-98}$ & 0.15 \\ 
8 & Low vs High category scales & 4664 & $\chi^2$ & 0.004 & 0.077 & 0.80 & 0.28 & $10^{-57}$ & 0.01 \\ 
9 & Gain vs. loss framing & 4925 & $\chi^2$ & 0.007 & 0.029 & 0.84 & 0.68 & $10^{-91}$ & 0.001 \\ 
10 & Sex differences in implicit math attitudes & 4558 & $t$ & 0.010 & 0.05 & 0.76 & 0.00 & $10^{-56}$ & 0.15 \\ 
11 & quote attribution & 4964 & $t$ & 0.260 & 0.063 & 0.56 & 0.00 & $10^{-36}$ & 0.23 \\ 
12 & sunk costs & 4967 & $t$ & 0.110 & 0.032 & 0.36 & 0.00 & $10^{-19}$ & 0.19 \\ 
13 & norm of reciprocity & 4928 & $\chi^2$ & 0.008 & 0.04 & 0.44 & 0.20 & $10^{-28}$ & 0.07 \\ 
14 & imagined contact & 4973 & $t$ & 0.110 & 0.034 & 0.16 &   0.00 & $10^{-7}$ & 0.34 \\ 
15 & Flag priming & 4896 & $t$ & 0.012 & 0.017 &   0.00 &   0.00 & 0.20 & 0.45 \\ 
16 & Currency priming & 4970 & $t$ & 0.016 & 0.021 &   0.00 &   0.00 & 0.40 & 0.49 \\ 
\hline
 	\end{tabular}
  	} 
\end{table}
 
 For every individual experiment, in this script we calculated the within-experiment sample variance $S^2$; for each experimental task we calculated the between-experiment sample variance $S_0^2$ (in terms of sample means $\overline{X}$ or slopes $\hat{M}$).  To assess the distributional model of significance against these experimental results, we used these sample variances to calculate the distributional significance estimate $p_{sig}$ for each experiment, and compared against the point-form significance $p$ for each experiment.

 \begin{figure*}[t!] 
 	\begin{center}
 		\scalebox{0.65}{\includegraphics*[viewport= 0 20  700 350]{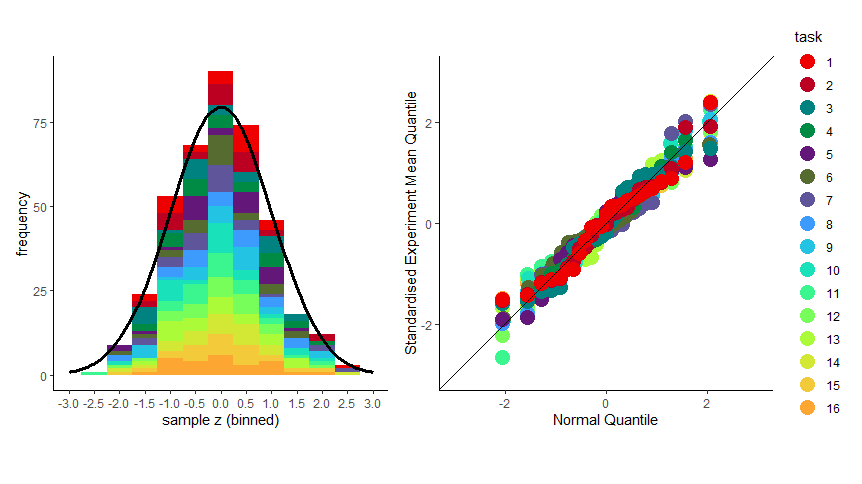}}
 	\end{center} 
 	\caption{ \label{fig:means_graph} \textbf{(Left)} Histogram of the standardised sample means, $z=(\overline{X}_i -\overline{X}_*)/S_0$, for US sites for each task in the Many Labs 1 replication dataset, in bins of size $0.5$ with colour representing task. The line shows the standard Normal distribution $ \mathcal{N}\left(0,1\right)$, scaled by  bin size $\times$  number of observations to match the histogram scale. \textbf{(Right)} QQ plot showing quantiles of the standardised sample means, $z=(\overline{X}_i -\overline{X}_*)/S_0$, for US sites for each task against quantiles of the standard Normal distribution. }
 \end{figure*}

 To assess the distributional model of replication, we grouped all experiments for a given task into `predictor-target' pairs and used the results of the predictor experiment to calculate the probability of replication $p_{rep}$ for that particular target experiment (and  a given value $\alpha$).  We then compared the predicted replication probability for each target experiment against the observed rate of successful replication in those target experiments.

 \subsection{Distribution of means across experiments} 
 To test the assumption that experimental means vary between experiments following a Normal distribution, for each task we calculate $\overline{X}_*$ (the average of sample means $\overline{X}_i$ for each replication of that task; for correlation tests we take $\overline{X}_i = \hat{M}_i$) and $S^2_0$ (the sample estimate of the variance of means or slopes $\mu_i$ around the overall mean $\mu_0$), and  normalise to $z= (\overline{X} - \overline{X}_* )/S_0$.

 Figure \ref{fig:means_graph}(left)  plots a histogram of these $z$ values  and shows consistent  variance in experimental means $\overline{X}$ for all tasks, with $z$ scores for means following the standard Normal distribution.  Figure \ref{fig:means_graph}(right) plots the quantiles of these  $z$ values for each individual task against quantiles of the standard Normal distribution, confirming that individual experimental means for each task follow this distribution.  Shapiro-Wilk tests  showed no significant evidence against the Normal distribution for all but $1$ task (point-form $p > 0.05$ for all but $1$ task; note that since the specific model here is one where experimental means $\mu_i$ are normally distributed around a point-form mean $\mu_0$, the point-form test for normality is appropriate in this case).  Together these results support the distributional model's assumption that means of experimental replications vary randomly following an approximately Normal distribution.  
 
 \subsection{Comparing point-form and distributional significance}
  Statistical significance  in this dataset was originally assessed in a two-sided test relative to the point-form $p$ at $\alpha = 0.05$, with a given experimental result counted as significant when it was both significant in this two-sided test and where the effect was the same direction as that seen in the originally published experiment (equivalent to a one-sided test in the original direction with $\alpha=0.025$). We follow this definition for distributional significance.  

As an initial check we used numerical integration methods to calculate, for each experiment, the theoretical value $p_{sig}(t|\hat{b})$ given that experiment's $t$ value and the $\hat{b}$ estimate for that task, and compared against the closed-form expression $p_{sig}$ for that experiment.    $p_{sig}(t|\hat{b})$ and $p_{sig}$ values were almost identical (correlation $r=0.99$), supporting the closed-form approximations, and so we used only the closed-form approximation $p_{sig}$ in our analysis.

To compare point-form and distributional measures of significance ($p$ and $p_{sig}$) we calculate, for each task, the $p$ value for results from each US site and record the percentage of sites which have $p \leq 0.05$.   Taking  $\hat{b}=S_0^2/S^2$ for each task and experiment we similarly calculate $p_{sig}$ values  for each site and record the percentage of sites with $p_{sig} \leq 0.05$.
Table $1$  compares, for US sites, the percentage of successful replications of a given task at the $\alpha=0.05$ level  under  point-form ($p$) and distributional ($p_{sig}$) tests, and also reports the $p$ and $p_{sig}$ values for the aggregated data (across all US sites) for each task.  Point-form $p$ values for aggregated data are unreasonably extreme: $4$ tasks had  $p <10^{-325}$ (the computational minimum); a task with only   $16\%$  successful replications relative to the point-form $p$ had  aggregate significance of  $p < 10^{-7}$.    Distributional significance values $p_{sig}$ were orders of magnitude more conservative, both in terms of proportion of successful replications and in the aggregate measure.

\begin{figure*}[t!] 
	\begin{center}
		\scalebox{0.6}{\includegraphics*[viewport= 10 10 820 400]{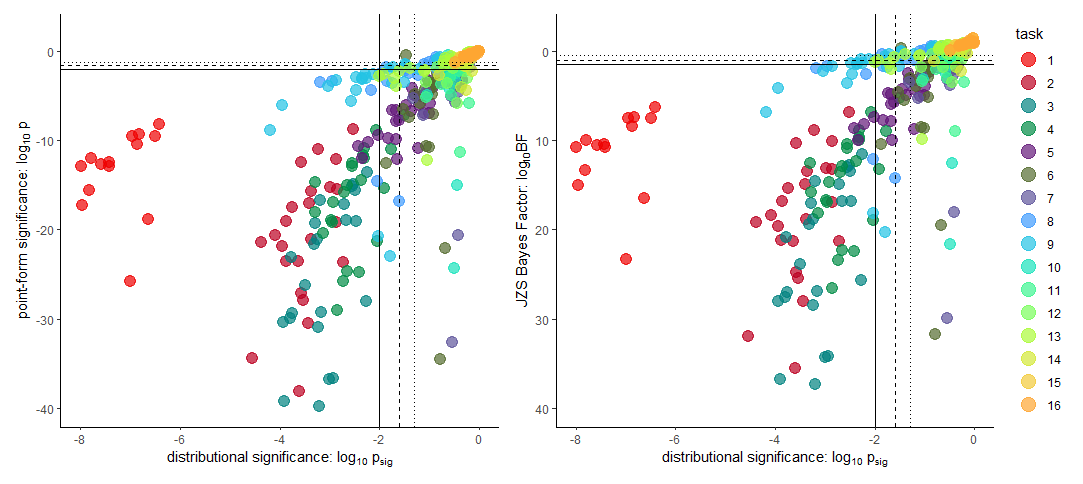}}
	\end{center} 
	\caption{ \label{fig:dist_vs_point} \textbf{(Left)} Scatterplot of $\log_{10}$  distributional $p_{sig}$ against $\log_{10}$ point-form $p$  for all experiments in all tasks  in the Many Labs 1 replication dataset for US sites (note the difference in scales).   Lines indicate statistical significance at levels  $\alpha=0.05$ (dotted), $\alpha=0.025$ (dashed) and $\alpha=0.01$ (solid).   \textbf{(Right)} Analogous scatterplot of $\log_{10}$  distributional $p_{sig}$ against $\log_{10}$ Bayes Factor $BF_{10}$. Horizontal lines indicate Bayes Factor levels of $3$ (dotted: $BF_{10} \geq 3$ being typically interpreted as moderate evidence against the null), $10$ (dashed: strong evidence against the null) and $30$ (solid:extremely strong evidence against the null).  Points for $9$ experiments, all with $p < 10^{-40}$ ($BF_{10} > 10^{40}$) and ranging as low as $p < 10^{-325}$ ($BF_{10} > 10^{325}$), are not shown.  Note that $\log_{10} p$ and  $\log_{10} BF$ values are identical modulo difference in scale (a $r=0.9999$ correlation between $\log_{10} p$ and  $\log_{10} BF$ across experiments).  }
\end{figure*}

We also compared $p_{sig}$ and point-form $p$ values for each individual result on a $\log_{10}$ scale (see Figure \ref{fig:dist_vs_point}(Left)). There are many extremely low $p$ values, with  $37\%$ of experiments having $p < 10^{-5}$ and  $22\%$ of experiments having $p < 10^{-15}$.   Distributional $p_{sig}$ values were again orders of magnitude more conservative ($94\%$ greater than $10^{-5}$ and all greater than  $10^{-10}$).   Many results were significant under point-form  but not distributional tests:  $40\%$ of  results with $p<0.05$ had $p_{sig} > 0.05$ and $11\%$ of results with $p < 10^{-5}$  had $p_{sig}  > 0.05$.  Even apparently highly significant results (in point-form tests) can turn out to be consistent with the null hypothesis when between-experiment variation is taken into account.

\subsection{Comparing predicted and observed replication probabilities}
To assess the distributional model of replication, for each of the $16$ tasks we generated every possible pair of experiments from two different US sites and  calculated the replication probability $p_{rep}$ for the second or target experiment in each pair based on the  result $t$ in the first or predictor experiment.  This calculation used the  between-experiment variance of means for the task overall, $S_0^2$, the within-experiment variance in the predictor experiment $S^2$ and the sample size of the predictor experiment $N$, and, given these values, generating the probability of replication for an experiment with the sample size $N_r$ of the paired target experiment.  This calculation assumed $c=1$ and so did not make use of the sample variance of the target experiment (and so represents the calculation a researcher would carry out when they have observed a surprising result in an experiment with sample size $N$ and within-experiment variance $S^2$, and are estimating the probability of replicating that result in a planned replication with sample size $N_r$ but unknown within-experiment variance).   We calculated this replication probability for a range of commonly used significance levels $\alpha = 0.05,0.025,0.01, 0.005,0.001$. For each target experiment and each level $\alpha$ we compared this predicted replication probability against a dichotomous successful replication variable ($p_{sig} \leq \alpha$) which indicated whether the target experiment did, in fact, replicate successfully at level $\alpha$.

As an initial check we calculated, for each pair of experiments in a given task, both the closed-form approximation $p_{rep}$ and the double-integral expression $p_{rep}(t|\hat{b})$.  As before, these were almost identical (correlation $r=0.94$) supporting the closed-form approximations, and so we primarily make use of the closed-form approximation $p_{sig}$ in our analysis (Figure $3$ gives a visual comparison of $p_{rep}$ and  $p_{rep}(t|\hat{b})$ values).

 \begin{table}[t!]
	\caption{\label{tab:replication_alpha}   Correlation between predicted and observed replication rates at each level of $\alpha$, with distributional significance $\hat{p}_{sig}(B=1)$ and replication probability $\hat{p}_{rep}(B=1,\alpha=0.05)$ for these correlations.  The table also gives the relevant parameters characterising each correlation: slope, predictor sum-of-squares $Q$ and response variation or error $S^2$.  }
	\centering
	\begin{tabular}{lcccrcrrr}
		\hline
		$\alpha$ & $N$ & $\%$ significant  & correlation $r$ for  & $\hat{p}_{sig}$ & $\hat{p}_{rep}$ & slope & $Q$ & $S^2$ \\
		& &   ($p_{sig} < \alpha$) & $p_{rep}$ vs ($p_{sig} < \alpha$)  & for $r$ & for $r$  & &\\ 
  \hline
0.05 &    9600 &  48 & 0.76 & 0.002 &    1 & 0.95 &   1534 & 0.11 \\ 
0.025 &    9600 &  40 & 0.78 & 0.0007 &    1 & 0.97 & 1499.8 & 0.09 \\ 
0.01 &    9600 &  34 & 0.80 & 0.0002 &    1 & 1.01 & 1367.5 & 0.08 \\ 
0.005 &    9600 &  30 & 0.82 & 0.00003 &    1 & 1.05 &   1224 & 0.07 \\ 
0.001 &    9600 &  20 & 0.78 & 0.00002 &    1 & 1.04 & 846.18 & 0.06 \\ 
all &   48000 &  34 & 0.80 & 0.0002 &    1 & 1.01 & 6738.7 & 0.083 \\ 
\hline
 \hline
	\end{tabular}
\end{table}

 Correlation between $p_{rep}$ values and dichotomous successful replication values ($p_{sig} \leq \alpha$)  was $r=0.80$ across the entire set with a slope of $1.01$. Point form $p$ for these correlations between predicted and observed replication was less than $10^{-325}$ (the computational minimum) in all cases. Table $2$ gives the correlation, significance and replication values for each level of $\alpha$ and overall.   The table also gives the necessary statistics for future calculations of between-experiment variance using these results (the sample slope, sum of squares of the predictor variable, and sample variance or error of the response variable).

\begin{figure*}[t!] 
	\begin{center}
		\scalebox{0.7}{\includegraphics*[viewport= 0 0 800 400]{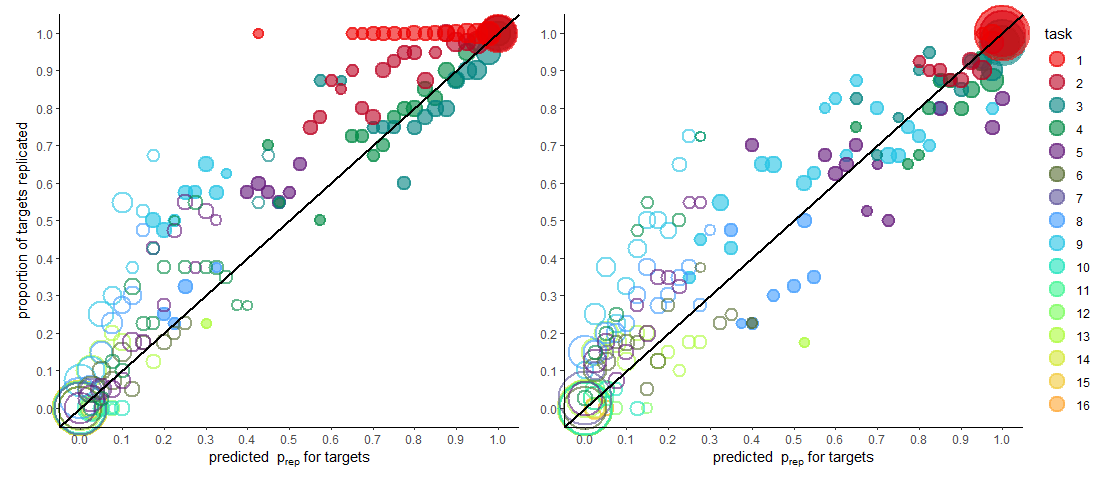}}
	\end{center} 
	\caption{ \label{fig:replication} Predicted vs observed replication rate (for $\alpha =0.05,0.025,0.01,0.005,0.001$) for target experiments in the Many Labs 1 replication dataset for US sites, grouped by $p_{rep}$ in bins of size $1/40$.   Bubble size represents the number of predictor-target pairs in each bin (sizes range from $40$ pairs for the smallest to $2800$ for the largest bubble).  Hollow bubbles contain only non-significant and solid bubbles  only significant predictor experiments.  The left graph shows results for the double-integral $p_{rep}(t|\hat{b})$ replication estimate, the right for the closed-form $p_{rep}$ estimate.  Correlation between predicted and observed replication rates for bubbles was high for $p_{rep}(t|\hat{b})$ ($r=0.91,\hat{p}_{sig} < 10^{-11}$) and for the approximation $p_{rep}$ ($r=0.81,\hat{p}_{sig} < 10^{-7}$, $\hat{p}_{rep}=1.0$  with $\alpha=0.05$ and $B=1$ for both expressions). }
\end{figure*}

To illustrate the relationship visually, we combined the sets of predictor-target pairs for these  values of $\alpha$ into a single collection, and grouped those pairs into bins of size $1/40$ by level of $p_{rep}$ and task: each bin contained a set of predictor-target pairs involving the same task  with approximately the same  replication probability $p_{rep}$ produced from the predictor experiment for the target (for the value $\alpha$ initially applied to that pair).  To distinguish between predictions from statistically significant and non-significant predictor experiments, we carried out this grouping process once for all pairs for which the predictor experiment was significant at $p_{sig} \leq \alpha$, and a second time for all pairs where the predictor experiment was non-significant.  To estimate the true rate of successful replication for target experiments in each bin we calculated the proportion of pairs in each bin for which the target experiment was significant ($p_{sig} \leq \alpha$ held for the pair's assigned value of $\alpha$).  We excluded bins containing less than $40$ predictor-target pairs, to allow a relatively accurate estimate of the actual replication rate for experiments in each bin (and to produce a matched level of resolution for predicted and observed replication rates).      Figure \ref{fig:replication}  shows a bubbleplot of predicted and observed replication rates for the double integral $p_{rep}(t|b)$ and for  $p_{rep}$.  As the figure shows there was a reliable linear relationship between observed and predicted replication rates across bins. 

One  concern is that the observed relationship between predicted and observed replication probabilities could simply be an artefact of the differences in  significance rates across different tasks. To see the problem,  assume that a statistically significant experiment necessarily produces a high predicted replication probability, while a non-significant experiment necessarily produces a low predicted replication probability.  In this situation a task where all or almost all experiments were statistically significant would produce both a high predicted replication rate and (since every experiment is significant) a high observed replication rate (for the same reason), while  a task where all or almost all experiments were non-significant would produce both a low predicted replication rate and a low observed replication rate.  Figure \ref{fig:replication} shows that this artefact does not explain these results: predicted and observed replication probabilities were related at the centre of the probability scale, as well as at the boundaries.

\begin{figure*}[t!] 
	\begin{center}
		\scalebox{0.9}{\includegraphics*[viewport= 0 0  600 450]{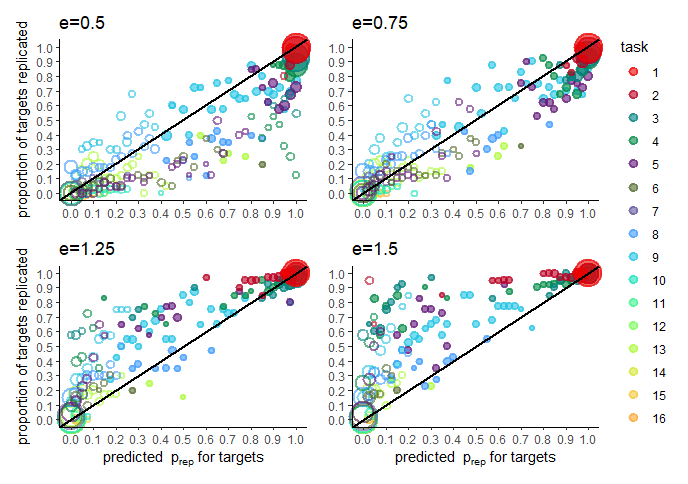}}
	\end{center} 
	\caption{ \label{fig:rep_scale_graphs} Sensitivity analysis.  Each graph shows the predicted vs observed replication rate (for $\alpha = 0.05,0.0.25,0.01,0.005,0.001$) for target experiments in the Many Labs 1 replication dataset for US sites as in Figure \ref{fig:replication}, but with between-experiment variation $S_0^2$ multiplied by scale $e$.   }
\end{figure*}

\subsection{Sensitivity to estimates of between-experiment variance}
The predicted probability of replication in the distributional model, $p_{rep}$, depends on an estimate of between-experiment variance.   In this section we ask how sensitive the accuracy of this predictive probability is to the between-variance estimate used in making the prediction.  To test this we re-ran our replication assessment on the US sites in the Many Labs $1$ dataset just as above, but with the between-experiment variance estimate $\hat{\sigma_0^2}$ for each task multiplied by a scaling value $e$.  In the first sensitivity analysis we set $e=0.5$ for all tasks (simulating a situation where the between-experiment variance used in calculating $p_{rep}$  is underestimated by $50\%$ relative to the actual experimental value for each task); in subsequent analyses we set $e$ to $0.75$, $1.25$ and $1.5$ (calculating $p_{rep}$ with between-experiment variance estimates of $75\%$, $125\%$ or $150\%$ of the actual value).

Correlations between $p_{rep}$ values and dichotomous successful replication values for target experiments, combined across all $\alpha$ levels, were $r= 0.75$ or higher for all values of $e$, indicating that even with significant over- or underestimation of between-experiment variance, $p_{rep}$ reliably predicts replication rates.   Figure \ref{fig:rep_scale_graphs} shows the predicted/observed replication rates produced in these analyses; correlation  between predicted and observed replication rates for bubbles was above $r=0.75$ for all values of $e$ ($\hat{p}_{sig} < 10^{-5}, \hat{p}_{rep}=1$, with $\alpha=0.05$ and $B=1$).  As the figure demonstrates, replication probability tended to be underestimated when between-experiment variance was underestimated ($e < 1$) but overestimated when between-experiment variance was overestimated ($e > 1$), with the degree of under/overestimation in replication increasing with the degree of under/overestimation in between-experiment variance.  This pattern follows from the relationship between type $1$ and $2$ error rates and the value of $b$.  There was relatively little difference in prediction accuracy for $e=0.75$ and $e=1.25$, suggesting that the $p_{rep}$ is not overly sensitive to error in estimates of between-experiment variance.   
  
\section{Comparing with Bayesian approaches}
This distributional model  mixes `frequentist-style' hypothesis testing with `Bayesian-style' updating and posterior prediction for estimating replication probabilities.    Advocates of the Bayesian approach may consider that we are `teetering on the edge of Bayes already' and ask why we do not go `all the way', by abandoning tests of statistical significance and moving to alternative Bayesian approaches.   In this section we address this possibility by comparing the distributional model against  the best-known and most commonly used Bayesian alternative to significance testing, the Bayesian or JZS $t$ test \citep{rouder2009Bayesian}.  

\subsection{JZS Bayesian $t$ test}
Given some observed data $\overline{X}$ on a parameter of interest the Bayes Factor $BF_{10}$ comparing two hypotheses $H_1$ and $H_0$ is the ratio of the  marginal likelihood of $\overline{X}$ given $H_1$  to the marginal likelihood  of  $\overline{X}$ given $H_0$:
\[BF_{10} = \frac{P(\overline{X}|H_1)}{P(\overline{X}|H_0)}\]
 Where a (point-form or distributional) $p$-value only gives evidence against the null hypothesis $H_0$, this measure $BF_{10}$ is taken to give evidence against $H_0$ and in favour of $H_1$: the higher $BF_{10}$, the more $\overline{X}$ gives Bayesian evidence in favour of $H_1$, while the higher its inverse $BF_{01}=1/BF_{10}$, the more $\overline{X}$ gives evidence in favour of $H_0$.

To use a Bayes Factor approach to assess evidence for or against a null hypothesis, the JZS Bayesian $t$ test \citep{rouder2009Bayesian} takes the parameter of interest to be the sample mean $\overline{X}$ and  takes $H_0$ to be the standard point-form null 
\[ H_0:  \overline{X} \sim \mathcal{N}(0,\sigma^2/N) \]
with unknown variance $\sigma$, so the probability of $\overline{X}$ given $H_0$ is the density of this distribution at $\overline{X}$,
\[ p(\overline{X}|H_0)= \int_0^{\infty}  \mathcal{N}(\overline{X}|0,\sigma^2) f_{\sigma}(\sigma) \mathrm{d}\sigma \]
where $f_{\sigma}$ is the density of the assumed prior for $\sigma$ (which in the JZS test is the Jeffreys prior). 
The alternative hypothesis $H_1$ is that the unknown mean $\mu$ itself follows a Normal distribution $\mu \sim   \mathcal{N}(0,\sigma_{\mu}^2)$, so that the parameter of interest follows the distribution 
\[ H_1:  \overline{X} \sim \mathcal{N}(0, \sigma^2/N + \sigma_{\mu}^2) \]
(identical to our distributional null model) so that
\[ p(\overline{X}|H_1)= \int_0^{\infty} \int_0^{\infty}  \mathcal{N}(\overline{X}|0, \sigma^2/N + \sigma_{\mu}^2)  f_{\sigma}(\sigma) f_{\sigma_\mu}(\sigma_\mu)\, \mathrm{d}\sigma\, \mathrm{d}\sigma_\mu \]
where $f_{\sigma_\mu}$ is the prior density of $\sigma_\mu$.\footnote{Note that this distribution is typically presented in term of a reparameterisation to effect size $\delta$ with unknown variance $\sigma_\delta^2$, where $\sigma_\delta^2$ is assumed to have an inverse $\chi^2(1)$ prior.  This reparmeterisation has no effect, of course, on the associated density: we retain the expression in terms of $\mu$ to allow readers to compare with our analogous expression for the distributional null.}

The JZS Bayesian $t$ test is implemented in the R BayesFactor package via the function `ttest.tstat' \citep{BayesFactor2021}. 
To compare this approach against our distributional null proposal, we used this package to calculate $BF_{10}$ for every experiment in the Many-Labs dataset, and compared these $BF_{10}$ values against the distributional significance   $p_{sig}$ for that result on a $\log_{10}$ scale (Figure  \ref{fig:dist_vs_point}(Right)).  This comparison gave results almost identical to those seen in the earlier point-form $p$ comparison.    There were many extremely high $BF_{10}$ values, with  $33\%$ of experiments having $BF_{10} > 10^{5}$ and  $17\%$ of experiments having $BF_{10} > 10^{15}$.   Distributional $p_{sig}$ values were again orders of magnitude more conservative.   Many results gave strong or very strong evidence against the null in the Bayesian test while not reaching the minimum level of significance in the distributional test:  $20\%$ of results with $BF_{10} > 30$ (very strong evidence against the null)  had $p_{sig}  > 0.05$.  Even results with apparently very strong evidence against the null (in the Bayesian $t$ test) turn out to be consistent with the null hypothesis  when between-experiment variation is taken into account.

 The left and right scatterplots in Figure \ref{fig:dist_vs_point} are extremely similar, suggesting that the logs of the point-form $p$ and the Bayes Factor $BF_{10}$ are related; or more precisely (since Figure \ref{fig:dist_vs_point}(Left) has a negative base-10 log scale on the $y$ axis) that $\log_{10} (BF_{10})$ and $\log_{10} (1/p)$ are related.  A comparison of values of  $\log_{10} (BF_{10})$ versus $\log_{10}(1/p)$ across experiments and tasks in the Many Labs dataset and confirms that these measures have an almost perfect linear relationship, with a correlation of $r=0.9999$ and with the line of best fit being $ \log_{10} (BF_{10}) = 0.99  \log_{10}(1/p) - 1.59$.   This means that the point-form and the Bayesian $t$ test  convey essentially the same information, at least in this dataset.   This close relationship between the point-form $p$ value and the Bayesian $t$ test measure is not surprising: it has long been observed empirically that Bayes Factor measures involving the point-form null and the point-form $p$-value measure are strongly correlated \citep[e.g.][]{wetzels2011statistical}  and that decisions based on these two measures are often equivalent  \citep[to quote Jeffreys: ``As a matter of fact I have applied my significance tests to numerous applications that have also been worked out by Fisher’s, and have not yet found a disagreement in the actual decisions reached''; cited in][]{ly2016harold}.  Given this relationship it seems clear that the various problems with point-form hypothesis testing  (failures of replication for significant results,  the fact that the point-form null hypotheses are always false, the effects of sample size on significance, and so on) will also apply to Bayesian approaches that involve the point-form null: we argue that these problems can be addressed, not by moving from a Frequentist to a Bayesian position, but instead by moving from point-form to distributional nulls.

\section{Conclusions}
The above results give evidence in favour of the distributional approach to significance testing, supporting its assumptions about Normally-distributed between-experiment variance of means, showing that many results that appear statistically significant under the point-form approach are identified as consistent with with random (between-experiment) variation in the distributional approach, and showing that the predicted replication probabilities generated for target experiments in the distributional approach are closely related to the actual observed probability of successful replication in those experiments, at least in the Many Labs 1 dataset.   

 We believe this distributional approach  has a number of advantages over  standard point-form significance testing.  First, this distributional approach more accurately reflects the various random factors that do, in fact, influence all experimental outcomes and so gives a more accurate estimate of the degree to which a given outcome is unexpected or surprising under a null model of random variation but no real effect.   Second, this approach addresses well-known problems with sample size that arise in the standard approach, where the probability of a significant result rises with sample size irrespective of any real effect.  This issue is particularly important in research involving hypothesis testing on large data sets (increasingly common in the `big data' era), and so the distributional approach should be useful in this form of research.
   Third, because the distributional model includes between-experiment variation in statistical analysis, it gives a natural account for the rate at which significant results will replicate in future experiments (an account which the point-form model cannot provide, because that model does not allow for cross-experiment variation in the experimental mean).
  Finally, this approach also gives a useful perspective on  commonly used measures of the statistical power of an experiment to detect effects, showing that in the presence of between-experiment variation some effects simply cannot be detected reliably, no matter how large the sample.

 We expect a number of objections to our argument for a distributional approach to hypothesis testing.  The first objection concerns the use of a  distributional representation of the null hypothesis. `These distributional nulls are just  Bayesian priors in another form' we imagine the objection goes, `and Bayesian priors are subjective measures of belief, not objective probability estimates.  Subjective beliefs cannot enter into objective frequentist hypothesis testing'.  
 
 It is true that our distributional nulls have a mathematical form that is identical to a Bayesian prior.  It is not true, however, that these distributional nulls \textit{necessarily} represent subjective measures of belief:  in a Frequentist interpretation, they can be taken to estimate between-experiment variance.  Indeed, these distributional nulls play the same  role that point-form nulls play in standard significance testing: for a given (distributional or point-form) null hypothesis  we say `result $t$ would have a low probability of occurrence if null hypothesis were true, and so result $t$ is unlikely to be simply a consequence of random processes'.  Such  assertions do not require or reflect any subjective belief in the null hypothesis.  To put this response another way: since the logic of significance testing is independent of the form of null hypothesis being used, our use of distributional rather than point nulls does not change the objective nature of such hypothesis testing.
 
 A second objection concerns statistical testing against distributional null hypotheses characterised by the parameter $b$ (representing  the between-experiment to within-experiment variance ratio).  `Researchers can adjust the distribution of this parameter  $b$ until they find a null hypothesis against which their observed results are statistically significant' we imagine the objection goes. `But this is simply a form of data-dredging or p-hacking: an attempt to find patterns in data that can be presented as statistically significant when in fact there is no real underlying effect'.  
 
  Our response here is to point out that this value $b$ is not picked arbitrarily: instead,  $b$ is estimated from experimental data on replications (and we give unbiased estimators for that situation).  More generally, even if researchers were arbitrarily picking values for this parameter $b$, the use of distributional testing systematically reduces the occurrence of statistically significant results, relative to the point-form null.  This is because  the rejection region for a null hypothesis falls monotonically with the value of this ratio $b$: and so the maximum rejection region (and so the greatest chance of a statistically significant result) arises with the point-form null.    A better characterisation of this process of testing against distributional nulls is one where we attempt to \textit{reduce} the chance of finding statistically significant results: where we are conservative in results as statistically significant, taking into account both between-experiment and within-experiment variance.
 
 A third possible objection concerns the availability of alternative hypothesis-testing methods, such as the Bayesian $t$ test (commonly put forward as a replacement for the significance testing approach).  `While significance testing was appropriate in the last century', we imagine the objection goes, `today we have better statistical methods; we don't need to test for significance'.   Our response here is to note that measures of significance in the point-form $t$ test and measures of evidence in the JZS Bayesian $t$ test  appear to be almost exactly equivalent in order of magnitude, at least in the Many Labs $1$ data, and so replacing significance testing against a point-form null with this form of Bayesian $t$ test may not, therefore, represent a meaningful advance.  
 
 While useful, our distributional approach is subject to various caveats.  First, our approach assumes normally distributed variation or error both within and between experiments.  This assumption is reasonable for simple experiments involving one or two variables ($t$ tests, tests of correlation or contingency) but does not  hold in more complex experimental designs involving multiple variables, such as ANOVAs, multiple regression analyses, and mixed models.   This is because the internal structure in such designs can cause random between-experiment differences to have systematic non-normal effects on between-experiment variability.  Applying our distributional approach to between-experiment variation in this more complex setting would be challenging, requiring a statistical model of between-experiment variance in each component of the experimental design and how that variance affects the overall between-experiment distribution of results.
  
Second, while our distributional approach allows us to estimate the probability of replication of given experimental results,  it is not a `magic bullet' that resolves all problems of replication.  In particular, our approach does not address the important problem of publication bias.  The distributional approach works best with some estimate for the between-experiment variability of means for a given experimental task,  $S^2_0$.  This can only be estimated from available (that is, published) data on experimental means.  If this set of available means is restricted in some way as a consequence of selective publication (if, for example, published means tend to be further from $0$ than unpublished means), this will lead to a systematic underestimation of between-experiment variation.    This problem of publication bias does not apply to the Many Labs $1$ dataset results: while the experimental tasks examined in that dataset were selected because they investigate well-known classic and contemporary effects in psychology (and so the original papers describing those effects were subject to some degree of publication bias), the results from all $36$ replications of those effects were `published' irrespective of their experimental means.  This problem clearly does apply, however, if between-experiment variability is estimated from a set of individually published experimental results.  This problem of publication bias is best addressed by continuing replication studies in various fields.  Such studies are additionally important because they will give useful estimates of the degree of between-experiment variation in experiments of different types, allowing researchers to assess the approximate probability of replication of new results in similar experiments.

  \pagebreak 
  
 \bibliographystyle{apalike}   
 \bibliography{new_references} 

\newcommand{\noop}[1]{}
\begin{thebibliography}{}

\bibitem[Amrhein and Greenland, 2018]{amrhein2018remove}
Amrhein, V. and Greenland, S. (2018).
\newblock {R}emove, {R}ather {T}han {R}edefine, {S}tatistical {S}ignificance.
\newblock {\em {N}ature {H}uman {B}ehaviour}, 2(1):4.

\bibitem[Berger and Pericchi, 1996]{berger1996intrinsic}
Berger, J.~O. and Pericchi, L.~R. (1996).
\newblock {T}he {I}ntrinsic {B}ayes {F}actor for {M}odel {S}election and
  {P}rediction.
\newblock {\em {J}ournal of the {A}merican {S}tatistical {A}ssociation},
  91(433):109--122.

\bibitem[Bishop, 2006]{bishop2006pattern}
Bishop, C.~M. (2006).
\newblock {\em {P}attern {R}ecognition and {M}achine {L}earning}.
\newblock Springer.

\bibitem[Camerer et~al., 2018]{camerer2018evaluating}
Camerer, C.~F., Dreber, A., Holzmeister, F., Ho, T.-H., Huber, J., Johannesson,
  M., Kirchler, M., Nave, G., Nosek, B.~A., Pfeiffer, T., et~al. (2018).
\newblock {E}valuating the {R}eplicability of {S}ocial {S}cience {E}xperiments
  in {N}ature and {S}cience {B}etween 2010 and 2015.
\newblock {\em {N}ature {H}uman {B}ehaviour}, 2(9):637--644.

\bibitem[Carver, 1978]{carver1978case}
Carver, R. (1978).
\newblock {T}he {C}ase {A}gainst {S}tatistical {S}ignificance {T}esting.
\newblock {\em {H}arvard {E}ducational {R}eview}, 48(3):378--399.

\bibitem[{CDF Collaboration} et~al., 2022]{cdf2022high}
{CDF Collaboration}, Aaltonen, T., Amerio, S., Amidei, D., Anastassov, A.,
  Annovi, A., Antos, J., Apollinari, G., Appel, J., Arisawa, T., et~al. (2022).
\newblock High-precision measurement of the {W} boson mass with the {CDF II}
  detector.
\newblock {\em Science}, 376(6589):170--176.

\bibitem[Chandler et~al., 2019]{chandler2019cochrane}
Chandler, J., Cumpston, M., Li, T., Page, M.~J., and Welch, V. (2019).
\newblock Cochrane handbook for systematic reviews of interventions.
\newblock {\em Hoboken: Wiley}.

\bibitem[Christensen, 2005]{christensen2005testing}
Christensen, R. (2005).
\newblock {T}esting {F}isher, {N}eyman, {P}earson, and {B}ayes.
\newblock {\em {T}he {A}merican {S}tatistician}, 59(2):121--126.

\bibitem[Cohen, 1992]{cohen1992power}
Cohen, J. (1992).
\newblock {A} {P}ower {P}rimer.
\newblock {\em {P}sychological {B}ulletin}, 112(1):155.

\bibitem[Cohen, 2013]{cohen2013statistical}
Cohen, J. (2013).
\newblock {\em {S}tatistical {P}ower {A}nalysis for the {B}ehavioral
  {S}ciences}.
\newblock Routledge.

\bibitem[Costello, 2022]{Costello_2022}
Costello, F. (2022).
\newblock Significance and replication with distributional null hypothesis
  testing.
\newblock \url{http://osf.io/ep86y}.

\bibitem[Costello and Watts, 2023]{costello2023underestimating}
Costello, F. and Watts, P. (2023).
\newblock Underestimating the uncertainty of aggregated results: the case of
  {W}-{B}oson mass.
\newblock {\em The European Physical Journal C}, 83(9):798.

\bibitem[Dahl et~al., 2019]{dahl2009xtable}
Dahl, D.~B., Scott, D., Roosen, C., Magnusson, A., and Swinton, J. (2019).
\newblock {\em {X}table: {E}xport {T}ables to {L}a{T}e{X} or {HTML}}.
\newblock R package version 1.8-4.

\bibitem[Dass and Berger, 2003]{dass2003unified}
Dass, S.~C. and Berger, J.~O. (2003).
\newblock Unified conditional frequentist and {B}ayesian testing of composite
  hypotheses.
\newblock {\em Scandinavian {J}ournal of {S}tatistics}, 30(1):193--210.

\bibitem[Diaconis et~al., 1991]{diaconis1991replication}
Diaconis, P. et~al. (1991).
\newblock [{R}eplication and {M}eta-{A}nalysis in {P}arapsychology]: {C}omment.
\newblock {\em {S}tatistical {S}cience}, 6(4):386--386.

\bibitem[Doros and Geier, 2005]{doros2005probability}
Doros, G. and Geier, A.~B. (2005).
\newblock {P}robability of {R}eplication {R}evisited: {C}omment on {{A}}n
  {A}lternative to {N}ull-{H}ypothesis {S}ignificance {T}ests.
\newblock {\em {P}sychological {S}cience}, 16(12):1005--1006.

\bibitem[Du et~al., 2023]{du2023explaining}
Du, X.~K., Li, Z., Wang, F., and Zhang, Y.~K. (2023).
\newblock Explaining the {CDF-II} {W}-boson mass anomaly in the
  {G}eorgi--{M}achacek extension models.
\newblock {\em The European Physical Journal C}, 83(2):1--13.

\bibitem[Endo and Mishima, 2022]{endo2022new}
Endo, M. and Mishima, S. (2022).
\newblock New physics interpretation of {W}-boson mass anomaly.
\newblock {\em Physical Review D}, 106(11):115005.

\bibitem[Fritz et~al., 2012]{fritz2012effect}
Fritz, C.~O., Morris, P.~E., and Richler, J.~J. (2012).
\newblock {E}ffect {S}ize {E}stimates: {C}urrent {U}se, {C}alculations, and
  {I}nterpretation.
\newblock {\em {J}ournal of {E}xperimental {P}sychology: {G}eneral}, 141(1):2.

\bibitem[Gabry et~al., 2019]{gabry2019visualization}
Gabry, J., Simpson, D., Vehtari, A., Betancourt, M., and Gelman, A. (2019).
\newblock Visualization in {B}ayesian workflow.
\newblock {\em Journal of the {R}oyal {S}tatistical {S}ociety {S}eries {A}:
  {S}tatistics in {S}ociety}, 182(2):389--402.

\bibitem[Ghorbani and Ghorbani, 2022]{ghorbani2022w}
Ghorbani, K. and Ghorbani, P. (2022).
\newblock W-boson mass anomaly from scale invariant 2hdm.
\newblock {\em Nuclear Physics B}, 984:115980.

\bibitem[G{\"o}nen et~al., 2005]{gonen2005bayesian}
G{\"o}nen, M., Johnson, W.~O., Lu, Y., and Westfall, P.~H. (2005).
\newblock {T}he {B}ayesian {T}wo-{S}ample {T} {T}est.
\newblock {\em {T}he {A}merican {S}tatistician}, 59(3):252--257.

\bibitem[Hedges, 1983]{hedges1983random}
Hedges, L.~V. (1983).
\newblock {A} {R}andom {E}ffects {M}odel for {E}ffect {S}izes.
\newblock {\em {P}sychological {B}ulletin}, 93(2):388.

\bibitem[Hunter, 1997]{hunter1997needed}
Hunter, J.~E. (1997).
\newblock {N}eeded: {A} {B}an on the {S}ignificance {T}est.
\newblock {\em {P}sychological {{S}}cience}, 8(1):3--7.

\bibitem[Iverson et~al., 2009a]{iverson2009prep}
Iverson, G.~J., Lee, M.~D., and Wagenmakers, E.-J. (2009a).
\newblock {P}rep {M}isestimates the {P}robability of {R}eplication.
\newblock {\em {P}sychonomic {B}ulletin \& {R}eview}, 16(2):424--429.

\bibitem[Iverson et~al., 2010a]{iverson2010random}
Iverson, G.~J., Lee, M.~D., and Wagenmakers, E.-J. (2010a).
\newblock {T}he {R}andom {E}ffects {P}rep {C}ontinues to {M}ispredict the
  {P}robability of {R}eplication.
\newblock {\em {P}sychonomic {B}ulletin \& {R}eview}, 17(2):270--272.

\bibitem[Iverson et~al., 2009b]{iverson2009fits}
Iverson, G.~J., Lee, M.~D., Zhang, S., and Wagenmakers, E.-J. (2009b).
\newblock {P}rep: {A}n {A}gony in {F}ive {F}its.
\newblock {\em {J}ournal of {M}athematical {P}sychology}, 53(4):195--202.

\bibitem[Iverson et~al., 2010b]{iverson2010model}
Iverson, G.~J., Wagenmakers, E.-J., and Lee, M.~D. (2010b).
\newblock {A} {M}odel-{A}veraging {A}pproach to {R}eplication: {T}he {C}ase of
  {P}rep.
\newblock {\em {P}sychological {M}ethods}, 15(2):172.

\bibitem[Kelley and Preacher, 2012]{kelley2012effect}
Kelley, K. and Preacher, K.~J. (2012).
\newblock {O}n {E}ffect {S}ize.
\newblock {\em {P}sychological {M}ethods}, 17(2):137.

\bibitem[Kenny and Judd, 2019]{kenny2019unappreciated}
Kenny, D.~A. and Judd, C.~M. (2019).
\newblock The unappreciated heterogeneity of effect sizes: Implications for
  power, precision, planning of research, and replication.
\newblock {\em Psychological methods}, 24(5):578.

\bibitem[Killeen, 2005a]{killeen2005alternative}
Killeen, P.~R. (2005a).
\newblock {A}n {A}lternative to {N}ull-{H}ypothesis {S}ignificance {T}ests.
\newblock {\em {P}sychological {S}cience}, 16(5):345--353.

\bibitem[Killeen, 2005b]{killeen2005replicability}
Killeen, P.~R. (2005b).
\newblock {R}eplicability, {C}onfidence, and {P}riors.
\newblock {\em {P}sychological {S}cience}, 16(12):1009--1012.

\bibitem[Killeen, 2007]{killeen2007replication}
Killeen, P.~R. (2007).
\newblock {R}eplication {S}tatistics.
\newblock {\em {B}est {P}ractices in {Q}uantitative {M}ethods}, pages 103--124.

\bibitem[Klein et~al., 2014]{klein2014investigating}
Klein, R.~A., Ratliff, K.~A., Vianello, M., Adams~Jr, R.~B., Bahn{\'\i}k,
  {\v{S}}., Bernstein, M.~J., Bocian, K., Brandt, M.~J., Brooks, B., Brumbaugh,
  C.~C., et~al. (2014).
\newblock {I}nvestigating {V}ariation in {R}eplicability.
\newblock {\em {S}ocial {{P}}sychology}, 45(3):142--152.

\bibitem[Klein et~al., 2018]{klein2018many}
Klein, R.~A., Vianello, M., Hasselman, F., Adams, B.~G., Adams~Jr, R.~B.,
  Alper, S., Aveyard, M., Axt, J.~R., Babalola, M.~T., Bahn{\'\i}k, {\v{S}}.,
  et~al. (2018).
\newblock {M}any {L}abs 2: {I}nvestigating {V}ariation in {R}eplicability
  {A}cross {S}amples and {S}ettings.
\newblock {\em {A}dvances in {M}ethods and {P}ractices in {P}sychological
  {S}cience}, 1(4):443--490.

\bibitem[Lipsey et~al., 2012]{lipsey2012translating}
Lipsey, M.~W., Puzio, K., Yun, C., Hebert, M.~A., Steinka-Fry, K., Cole, M.~W.,
  Roberts, M., Anthony, K.~S., and Busick, M.~D. (2012).
\newblock {T}ranslating the {S}tatistical {R}epresentation of the {E}ffects of
  {E}ducation {I}nterventions {I}nto {M}ore {R}eadily {I}nterpretable {F}orms.
\newblock {\em {N}ational {C}enter for {S}pecial {E}ducation {R}esearch}.

\bibitem[Ly et~al., 2016]{ly2016harold}
Ly, A., Verhagen, J., and Wagenmakers, E.-J. (2016).
\newblock {H}arold {J}effreys?s {D}efault {B}ayes {F}actor {H}ypothesis
  {T}ests: {E}xplanation, {E}xtension, and {A}pplication in {P}sychology.
\newblock {\em {J}ournal of {M}athematical {P}sychology}, 72:19--32.

\bibitem[Macdonald, 2005]{macdonald2005replication}
Macdonald, R.~R. (2005).
\newblock {W}hy {R}eplication {P}robabilities {D}epend on {P}rior {P}robability
  {D}istributions.
\newblock {\em {P}sychological {S}cience-{C}ambridge-}, 16(12):1007.

\bibitem[Mandel, 2012]{mandel2012statistical}
Mandel, J. (2012).
\newblock {\em {T}he {S}tatistical {A}nalysis of {E}xperimental {D}ata}.
\newblock Courier Corporation.

\bibitem[Maraun and Gabriel, 2010]{maraun2010killeen}
Maraun, M. and Gabriel, S. (2010).
\newblock {K}illeen's (2005) {P}rep {C}oefficient: {L}ogical and {M}athematical
  {P}roblems.
\newblock {\em {P}sychological {M}ethods}, 15(2):182.

\bibitem[McShane et~al., 2019]{BlakeleyAbandon2019}
McShane, B.~B., Gal, D., Gelman, A., Robert, C., and Tackett, J.~L. (2019).
\newblock {A}bandon {S}tatistical {S}ignificance.
\newblock {\em {T}he {A}merican {S}tatistician}, 73(sup1):235--245.

\bibitem[Meehl, 1990a]{meehl1990appraising}
Meehl, P.~E. (1990a).
\newblock {A}ppraising and {A}mending {T}heories: {T}he {S}trategy of
  {L}akatosian {D}efense and {T}wo {P}rinciples {T}hat {W}arrant {I}t.
\newblock {\em {P}sychological {I}nquiry}, 1(2):108--141.

\bibitem[Meehl, 1990b]{meehl1990summaries}
Meehl, P.~E. (1990b).
\newblock {W}hy {S}ummaries of {R}esearch on {P}sychological {T}heories {A}re
  {O}ften {U}ninterpretable.
\newblock {\em {P}sychological {R}eports}, 66(1):195--244.

\bibitem[Miller, 2009]{miller2009probability}
Miller, J. (2009).
\newblock {W}hat {I}s the {P}robability of {R}eplicating a {S}tatistically
  {S}ignificant {E}ffect?
\newblock {\em {P}sychonomic {B}ulletin \& {R}eview}, 16(4):617--640.

\bibitem[Miller and Schwarz, 2011]{miller2011aggregate}
Miller, J. and Schwarz, W. (2011).
\newblock {A}ggregate and {I}ndividual {R}eplication {P}robability {W}ithin an
  {E}xplicit {M}odel of the {R}esearch {P}rocess.
\newblock {\em {P}sychological {M}ethods}, 16(3):337.

\bibitem[Morey and Rouder, 2021]{BayesFactor2021}
Morey, R.~D. and Rouder, J.~N. (2021).
\newblock {\em {B}ayes{F}actor: {C}omputation of {B}ayes {F}actors for {C}ommon
  {D}esigns}.
\newblock R package version 0.9.12-4.3.

\bibitem[Murphy, 2007]{murphy2007}
Murphy, K.~P. (2007).
\newblock {C}onjugate {B}ayesian {A}nalysis of the {G}aussian {D}istribution.
\newblock Technical report, University of British Columbia.

\bibitem[{Open Science Collaboration} et~al., 2015]{open2015estimating}
{Open Science Collaboration} et~al. (2015).
\newblock {E}stimating the {R}eproducibility of {P}sychological {S}cience.
\newblock {\em {S}cience}, 349(6251):aac4716.

\bibitem[Orben and Lakens, 2020]{orben2020crud}
Orben, A. and Lakens, D. (2020).
\newblock {C}rud ({R}e) {D}efined.
\newblock {\em {A}dvances in {M}ethods and {P}ractices in {P}sychological
  {S}cience}, 3(2):238--247.

\bibitem[Pashler and Wagenmakers, 2012]{pashler2012editors}
Pashler, H. and Wagenmakers, E.-J. (2012).
\newblock {E}ditors' {I}ntroduction to the {S}pecial {S}ection on
  {R}eplicability in {P}sychological {S}cience: {A} {C}risis of {C}onfidence?
\newblock {\em {P}erspectives on {P}sychological {S}cience}, 7(6):528--530.

\bibitem[{R Core Team}, 2021]{Rmanual2021}
{R Core Team} (2021).
\newblock {\em {{R}}: {{A}} {{L}}anguage and {{E}}nvironment for
  {{S}}tatistical {{C}}omputing}.
\newblock R Foundation for Statistical Computing, Vienna, Austria.

\bibitem[Rouder et~al., 2009]{rouder2009Bayesian}
Rouder, J.~N., Speckman, P.~L., Sun, D., Morey, R.~D., and Iverson, G. (2009).
\newblock {B}ayesian {T} {T}ests for {A}ccepting and {R}ejecting the {N}ull
  {H}ypothesis.
\newblock {\em {P}sychonomic {B}ulletin \& {R}eview}, 16(2):225--237.

\bibitem[Sanabria and Killeen, 2007]{sanabria2007better}
Sanabria, F. and Killeen, P.~R. (2007).
\newblock {B}etter {S}tatistics for {B}etter {D}ecisions: {R}ejecting {N}ull
  {H}ypotheses {S}tatistical {T}ests in {F}avor of {R}eplication {S}tatistics.
\newblock {\em {P}sychology in the {S}chools}, 44(5):471--481.

\bibitem[Schad et~al., 2021]{schad2021toward}
Schad, D.~J., Betancourt, M., and Vasishth, S. (2021).
\newblock Toward a principled {B}ayesian workflow in cognitive science.
\newblock {\em Psychological {M}ethods}, 26(1):103.

\bibitem[Snedecor and Cochran, 1989]{snedecor1989statistical}
Snedecor, G.~W. and Cochran, W.~G. (1989).
\newblock {S}tatistical {M}ethods.

\bibitem[Tackett et~al., 2017]{tackett2017s}
Tackett, J.~L., Lilienfeld, S.~O., Patrick, C.~J., Johnson, S.~L., Krueger,
  R.~F., Miller, J.~D., Oltmanns, T.~F., and Shrout, P.~E. (2017).
\newblock It?s time to broaden the replicability conversation: Thoughts for and
  from clinical psychological science.
\newblock {\em Perspectives on Psychological Science}, 12(5):742--756.

\bibitem[Thompson, 1998]{thompson1998praise}
Thompson, B. (1998).
\newblock {I}n {P}raise of {B}rilliance: {W}here {T}hat {P}raise {R}eally
  {B}elongs.
\newblock {\em {A}merican {{P}}sychologist}, 53(7):799--800.

\bibitem[van~de Schoot et~al., 2021]{van2021bayesian}
van~de Schoot, R., Depaoli, S., King, R., Kramer, B., M{\"a}rtens, K., Tadesse,
  M.~G., Vannucci, M., Gelman, A., Veen, D., Willemsen, J., et~al. (2021).
\newblock {B}ayesian {S}tatistics and {M}odelling.
\newblock {\em {N}ature {R}eviews {M}ethods {P}rimers}, 1(1):1--26.

\bibitem[Veroniki et~al., 2016]{veroniki2016methods}
Veroniki, A.~A., Jackson, D., Viechtbauer, W., Bender, R., Bowden, J., Knapp,
  G., Kuss, O., Higgins, J.~P., Langan, D., and Salanti, G. (2016).
\newblock {M}ethods to {E}stimate the {B}etween-{S}tudy {V}ariance and {I}ts
  {U}ncertainty in {M}eta-{A}nalysis.
\newblock {\em {R}esearch {S}ynthesis {M}ethods}, 7(1):55--79.

\bibitem[Viechtbauer, 2005]{viechtbauer2005bias}
Viechtbauer, W. (2005).
\newblock {B}ias and {E}fficiency of {M}eta-{A}nalytic {V}ariance {E}stimators
  in the {R}andom-{E}ffects {M}odel.
\newblock {\em {J}ournal of {E}ducational and {B}ehavioral {S}tatistics},
  30(3):261--293.

\bibitem[Wagenmakers, 2007]{wagenmakers2007practical}
Wagenmakers, E.-J. (2007).
\newblock A practical solution to the pervasive problems of p values.
\newblock {\em Psychonomic bulletin \& review}, 14(5):779--804.

\bibitem[Wagenmakers et~al., 2018]{wagenmakers2018bayesian}
Wagenmakers, E.-J., Marsman, M., Jamil, T., Ly, A., Verhagen, J., Love, J.,
  Selker, R., Gronau, Q.~F., {\v{S}}m{\'\i}ra, M., Epskamp, S., et~al. (2018).
\newblock Bayesian inference for psychology. part i: Theoretical advantages and
  practical ramifications.
\newblock {\em Psychonomic bulletin \& review}, 25:35--57.

\bibitem[Wetzels et~al., 2011]{wetzels2011statistical}
Wetzels, R., Matzke, D., Lee, M.~D., Rouder, J.~N., Iverson, G.~J., and
  Wagenmakers, E.-J. (2011).
\newblock {S}tatistical {E}vidence in {E}xperimental {P}sychology: {A}n
  {E}mpirical {C}omparison {U}sing 855 {T} {T}ests.
\newblock {\em {P}erspectives on {P}sychological {S}cience}, 6(3):291--298.

\bibitem[Wickham, 2016]{ggplot2_2016}
Wickham, H. (2016).
\newblock {\em {G}gplot2: {E}legant {G}raphics for {D}ata {A}nalysis}.
\newblock Springer-Verlag New York.

\bibitem[Zhang and Feng, 2023]{zhang2023explaining}
Zhang, K.-Y. and Feng, W.-Z. (2023).
\newblock Explaining the {W} boson mass anomaly and dark matter with a {U}(1)
  dark sector.
\newblock {\em Chinese Physics C}, 47(2):023107.

\end{thebibliography}
    
   \pagebreak

   \section{Appendix 1}
 
Assuming $c=1$ and $N_r=N$ (an exact replication), for a given $t$ statistic and a
 between-experiment variance ratio $b$ we can rewrite $p_{rep}(t|b)$ as
 \begin{eqnarray*}
 	p_{rep}(t|b)&=&T_{\nu}\left(\frac{bN\left[\frac{|t|}{\sqrt{1+bN}}
 		-T_{\nu}^{-1}(1-\alpha/2)\right]-T_{\nu}^{-1}(1-\alpha/2)}{\sqrt{1+2bN}}\right)\\
 	&=&T_{\nu}\left(|t|\,f(bN)\right)
 \end{eqnarray*}
 where the function $f(z)$ is
 \begin{eqnarray*}
 	f(z)&=&\frac{1}{\sqrt{2z+1}}\left(\frac{z}{\sqrt{z+1}}-\frac{z+1}{\tau}\right)
 \end{eqnarray*}
 with $\tau=|t|/T_{\nu}^{-1}(1-\alpha/2)$ (which is positive for
 $\alpha<0.5$ and $t\neq 0$, both of which we assume).
 
 The derivative of this probability with respect to $b$ is
 \begin{eqnarray*}
 	\frac{\partial}{\partial b} p_{rep}(t|b)
 	&=&|t|NT_{\nu}'\left(|t|\,f(bN)\right)f'(bN),
 \end{eqnarray*}
 and since $T_{\nu}$ is a monotonically increasing function of its
 argument, this derivative will vanish only when $f'(bN)$ is zero.  A
 quick computation gives
 \begin{eqnarray*}
 	f'(z)&=&\frac{1}{(2z+1)^{3/2}}\left(\frac{3z+2}{2(z+1)^{3/2}}-\frac{z}{\tau}\right)
 \end{eqnarray*}
 and therefore $f$ -- and thus $p_{rep}(t|b)$ -- reaches its extremal
 value(s) at $z_0/N$, where
 \begin{eqnarray*}
 	f'(z_0)=\frac{1}{(2z_0+1)^{3/2}}
 	\left(\frac{3z_0+2}{2(z_0+1)^{3/2}}-\frac{z_0}{\tau}\right)&=&0.
 \end{eqnarray*}
 
 The above will be satisfied when
 \begin{eqnarray*}
 	\frac{3z_0+2}{2(z_0+1)^{3/2}}-\frac{z_0}{\tau}&=&0
 \end{eqnarray*}
 which can be rearranged to give the quintic equation
 \begin{eqnarray}
 \label{eq:quintic}
 z_0^5+3z_0^4+3z_0^3+\left(1-\frac{9\tau^2}{4}\right)z_0^2
 -3\tau^2z_0-\tau^2&=&0.\label{eq:quintic}
 \end{eqnarray}
 
 No matter what value $\tau$ takes, there is only one sign change
 between the coefficients of the polynomial on the left-hand side, so
 by Descartes' Rule of Signs, there is exactly one positive solution to
 this quintic; it can be shown that it is a maximum of $f(z)$, so we
 call it $z_{max}$.  It is this solution we use to obtain
 $b_{max}=z_{max}/N$, the value of $b$ which maximises $p_{rep}(t|b)$.
 
 To obtain a bound on $b_{max}$, we put $z_0 = \tau$ into the quintic
 polynomial in (\ref{eq:quintic}) and obtain
\begin{eqnarray*}
\tau^5+3\tau^4+3\tau^3+\left(1-\frac{9\tau^2}{4}\right)\tau^2
-3\tau^3-\tau^2 &=&\tau^5+\frac{\tau^4}{4}
\end{eqnarray*}
 which is necessarily positive.  Since the quintic in
 (\ref{eq:quintic}) has the negative value of $-\tau^2$ when $z_0=0$,
 it is zero for some value of $z_0$ between $0$ and $\tau$ and so
 $z_{max} \leq \tau$ and
 \[b_{max} \leq \frac{|t|}{N T_{\nu}^{-1}(1-\alpha/2)} =  \frac{|d|}{\sqrt{N} T_{\nu}^{-1}(1-\alpha/2)}\]
 where $d$ is the sample effect size.  Since this expression falls with $\sqrt{N}$, we see that $b_{max}$ is
 necessarily close to $0$ for all but very large effect sizes.

  \section{Appendix 2}
    
  Here we consider an experiment involving two  variables, a predictor variable $X$ and a response variable $Y$. For pairs $(X_i,Y_i)$ we assume that values of the predictor variable $X_i$ are fixed, and that values of the response variable  follow the distribution
  \begin{equation}
  Y_i =  \mathcal{N}\left (\mu\, X_i+a, \sigma^2 \right)
  \end{equation}
  for unknown slope $\mu$, offset $a$ and variance $\sigma^2$ (where $\sigma^2$ is the variance of random within-experiment error in the response variable; these are the standard assumptions of linear regression).
  
  We wish to test for independence of $X$ and $Y$.  In the point-form null model, this involves testing the hypothesis that the slope $\mu=0$. In the distributional model, this involves assuming that the slope $\mu$ varies randomly across experiments in a normal distribution $\mu \sim \mathcal{N}\left(\mu_0,b\sigma^2\right)$, and testing for independence involves  testing the hypothesis that  that $\mu_0=0$.
  
  Suppose we have $N$ pairs of estimates in our experiment. We take 
  \begin{equation*}
  Q = \sum_1^N (X_i- \overline{X})^2
  \end{equation*}
  to represent the sum-of-squares for the predictor variable $X$.    For a given pair of values  $(X_i,Y_i)$ we define $M_i = (X_i- \overline{X})Y_i$ as the estimate for the slope $\mu$ for that pair, and so take
  \begin{equation*}
  \overline{M} = \frac{\sum_1^N (X_i- \overline{X})Y_i}{ Q}
  \end{equation*}
  be our overall estimate for the slope $\mu$ .  Then the least-squares regression line for that experiment is
  \begin{equation*}
  \hat{Y_i} = \overline{Y} +  \overline{M} X_i
  \end{equation*}
  
  Within a single experiment the variable $\overline{M}$  is distributed around the unknown slope  $\mu$ in that experiment as 
  \begin{equation*}
  \mathcal{N}\left(\overline{M}| \mu,\frac{\sigma^2}{Q}\right)
  \end{equation*}
  Letting  
  \begin{equation*}
  S^2  = \frac{\sum_1^N (Y_i- \hat{Y_i})^2}{(N-2)}
  \end{equation*}
  be the mean squared error between observed and predicted values of the response variable $Y$ and, given the assumption of a linear relationship between $X$ and $Y$, the variable 
  \begin{equation*}
  \frac{(N-2) \, S^2}{\sigma^2}
  \end{equation*}
  has a $\chi^2$-distribution with $\nu=N-2$ degrees of freedom.   Then as before, under our distributional null hypothesis we have
  \begin{equation*}
  \overline{M} \sim \mathcal{N}\left( 0,\, (\sigma^2/Q)(1+ b Q) \right)
  \end{equation*}
  where $b=\sigma_0^2/\sigma^2$ (and so $b$ represents the ratio of between-experiment variance in the slope to  within-experiment variance in the response variable) and letting
  \begin{equation*}
  t = \frac{ \overline{M}}{S }\sqrt{Q}
  \end{equation*}
  the variable
  \begin{equation*}
  \frac{t}{ \sqrt{1+ b Q}}
  \end{equation*}
  has a $t$ distribution with $\nu$ degrees of freedom. These variables $\overline{M}$, $Q$, $S^2$ and $t/\sqrt{1+ b Q}$ for correlation thus have identical distributions and  relationships as the corresponding variables $\overline{X}$, $N$, $S^2$ and $t/\sqrt{1+ b N}$ for the t-test, and so estimates for significance, replication and between-experiment variance $S_0^2$ hold as given in the previous section (with these variable substitutions).  For example: given a set of $K$  replications of some similar correlation experiment, each with sample slope $\overline{M}_i$, predictor sum of squares $Q_i$,  mean squared response error $S_i^2$, and degrees of freedom $\nu_i$, this means that our estimate for the between-experiment variance of the slope is
  \begin{eqnarray*}
  	\langle  S_0^2 \rangle=	\frac{1}{K-1}\sum_{i=1}^K\left(\overline{M}_i-\overline{M}_0\right)^2
  	+\frac{1}{K}\sum_{i=1}^K\frac{\nu_i\, S_i^2}{Q_i(\nu_i-2)}
  \end{eqnarray*}
  where $\overline{M}_0$ is the mean of those sample slopes, and $ S_0^2$ follows an approximately $\chi^2$ distribution with degrees of freedom $\nu_0=K-1$.   Taking $ \hat{b} = S_0^2/S^2$ as our estimate for the ratio $b=\sigma^2_0/\sigma^2$ this means that we have a theoretical expression for significance of
  \[
  p_{sig}(r|\hat{b}) = \int_{0}^{\infty}2 T_{\nu}\left(-|t|/\sqrt{1+b\, \hat{b}\, Q}\right) f_{\nu,\nu_0}(b) \mathrm{d}b 
  \] 
  and a theoretical expression for replication (in a replication experiment with predictor sum-of-squares $Q_r$ and degrees of freedom $\nu_r$) of
  \[
  p_{rep}(r|\hat{b}) =   \int_{0}^{\infty} \int_{0}^{\infty}  T_{\nu_r}\left(  \frac{|t|\frac{ b\,\hat{b}\,\sqrt{Q \, Q_r} }{1+b\, Q}- T^{-1}_{\nu_r}\left(1-\alpha/2\right)\sqrt{c(1+b\,\hat{b}\,Q_r)}  }{ \sqrt{  c+b\,\hat{b}\,Q_r/(1+b\hat{b}\, Q) }  } \right)  f_{\nu,\nu_0}(b) f_{\nu,\nu_{r}}(c)  \mathrm{d}b \mathrm{d}c
  \]
From this we get closed-form expressions for significance and replication of
 \[
 p_{sig} = 2T_{\nu_0}\left(\frac{-|t|}{\sqrt{\hat{b}Q}}\right)
 \]
\[ p_{rep} =  T_{\nu_r}\left(\sqrt{\frac{\hat{b} Q Q_r}{Q+Q_r  } }\left[ \frac{-|t|}{\sqrt{\hat{b}Q}} -
T^{-1}_{\nu_r}\left(1-\alpha/2\right)\sqrt{ \frac{1}{\hat{b}Q}  + \frac{\nu_0}{\nu_0-2}}   \right] \right)
\]
and generic estimates of\[
\hat{p}_{sig} = 2T_{\nu}\left(-|t|/\sqrt{1+BQ}\right)
\] 
\[
\hat{p}_{rep}=\int_{0}^{\infty}  T_{\nu_r}\left(  \frac{|t|\frac{ B \sqrt{Q\, Q_r}}{1+BQ} - T^{-1}_{\nu_r}\left(1-\alpha/2\right)\sqrt{c(1+BQ_r)}  }{ \sqrt{  c+BQ_r/(1+BQ)  }  } \right) f_{\nu,\nu}(c) \mathrm{d}c
\]

  Note that in these correlation tests the sum of squares of the predictor variable $Q$ is the continuous-valued analog of the discrete sample size  $N$ in a $t$ test: where sample size gives the number of discrete measurements made  in a $t$ test,  predictor sum-of-squares gives the range of continuous measurements  in a test of correlation.    We've derived these expressions in a experimental design investigating how values of a response variable $Y$ are related to those of a predictor $X$ whose values are fixed by the experimenter.   However, this approach extends naturally to the situation where both $X$ and $Y$ are random variables.   To measure evidence for replication in this situation, however, it is necessary to select one of these variables as the predictor and the other as the response: an estimate for the probability of replication of a given result $r$ in an experiment where $X$ is the predictor and the $X$ sum of squares is assumed to have a particular controlled value will necessarily be different from the probability of replication of  that result in an experiment where $Y$ is the predictor the $Y$ sum of squares is assumed to have a particular controlled value .   
  
 We extend this approach to tests of independence in $2 \times 2$ contingency tables by considering an experiment involving $N$ pairs of outcomes from two dichotomous or binary variables, a predictor variable $X$ and a response variable $Y$, where we are interested in the degree of association between variable $X$ and variable $Y$.  The standard approach here is to count the number of times each pair of values for these variables occurs and apply Pearson's $\chi^2$-squared test to the resulting  contingency table, testing against the null hypothesis of no association between the two variables. Letting $n_{ij}$ represent the number of pairs where $X=i$ and $Y=j$, this test asks whether  the $\phi$ measure of association
  \begin{equation*}
  \phi = \frac{n_{11}n_{00}-n_{1 0 }n_{0 1}}{\sqrt{ (n_{1 1}+ n_{10})\, ( n_{11 }+ n_{01 })\, (n_{0 0}+n_{0 1})\,\, (n_{0 0}+n_{1 0})}}
  \end{equation*}
  is significantly different from $0$.  For $2 \times 2$ contingency table, however, this test is identical to a $Z$-test of proportions \citep{snedecor1989statistical}, and so does not take sample size into account.  This means that our analysis of replication (which depends on sample size or an analogous measure) cannot apply to the Pearson's $\chi^2$-squared test.  
  
  Note, however, that the correlation coefficient $r$ is given by
  \begin{equation*}
  r={\frac {N\sum X_{i}Y_{i}-\sum X_{i}\sum Y_{i}}{{\sqrt {N\sum X_{i}^{2}-(\sum X_{i})^{2}}}~{\sqrt {N\sum Y_{i}^{2}-(\sum Y_{i})^{2}}}}}
  \end{equation*}
  Since for two binary variables we have $N=n_{11}+n_{10}+n_{01}+n_{00}$, $\sum x_{i}y_{i}=n_{11}$, $\sum x_{i}=n_{10}+n_{11}$, and so on,  by substitution and rearrangement we see that $r=\phi$.
  Further, for large enough $N$ significance relative to the null hypothesis $\phi = 0$ in Pearson's $\chi^2$-squared test is equivalent to significance relative to the null hypothesis of $r=0$ in a test of linear correlation.    To use our distributional null model in experiments involving pairs of outcomes from dichotomous variables, we thus apply the correlation and linear regression approach described in the previous section to these dichotomous variables $X$ and $Y$.  
    \end{document}